\documentclass[journal=jctcce,manuscript=article]{achemso}

\usepackage{graphicx}
\usepackage{dcolumn}
\usepackage{bm}

\usepackage{lipsum}
\usepackage[table,xcdraw]{xcolor}
\usepackage{multirow}
\usepackage{soul}
\usepackage{simpler-wick}
\usepackage{chemformula}
\usepackage{xr}
\usepackage{hhline}
\usepackage{braket}
\usepackage[capitalize]{cleveref}
\usepackage{listings}
\usepackage{booktabs}
\usepackage{array}
\usepackage{siunitx}
\usepackage{threeparttable}
\usepackage{xcolor}
\usepackage{xspace}
\usepackage{pdfpages}

\newcommand{\bvec}[1]{\bm{\mathrm{#1}}}
\newcommand{\adj}{^{\dagger}}
\newcommand{\pz}{\Phi_0}
\newcommand{\barh}{\bar{H}}
\newcommand{\msp}{\mathcal{M}}
\newcommand{\ung}{_{\mathrm{u}}}
\newcommand{\ger}{_{\mathrm{g}}}
\newcommand{\statex}{\mathrm{X}}
\newcommand{\statea}{\mathrm{A}}
\newcommand{\stateb}{\mathrm{B}}
\newcommand{\sunit}{$E_\mathrm{h}^{-2}$\xspace}

\definecolor{goodorange}{RGB}{225,125,0}
\definecolor{goodgreen}{RGB}{0,125,0}
\definecolor{goodred}{RGB}{220,50,25}
\definecolor{goodblue}{RGB}{25,25,150}

\newcommand{\note}[2]{
\ifthenelse{\equal{#1}{F}}{
\colorbox{goodorange}{\textcolor{white}{\footnotesize \fontfamily{phv}\selectfont #1}}
    \textcolor{goodorange}{{\footnotesize \fontfamily{phv}\selectfont #2}}\xspace
}{}
\ifthenelse{\equal{#1}{S}}{
\colorbox{goodred}{\textcolor{white}{\footnotesize \fontfamily{phv}\selectfont #1}}
    \textcolor{goodred}{{\footnotesize \fontfamily{phv}\selectfont #2}}\xspace
}{}
\ifthenelse{\equal{#1}{Z}}{
\colorbox{goodgreen}{\textcolor{white}{\footnotesize \fontfamily{phv}\selectfont #1}}
    \textcolor{goodgreen}{{\footnotesize \fontfamily{phv}\selectfont #2}}\xspace
}{}
}
\definecolor{codegreen}{rgb}{0,0.6,0}
\definecolor{codegray}{rgb}{0.5,0.5,0.5}
\definecolor{codepurple}{rgb}{0.58,0,0.82}
\definecolor{backcolour}{rgb}{0.95,0.95,0.92}

\lstdefinestyle{mystyle}{
    backgroundcolor=\color{backcolour},   
    commentstyle=\color{codegreen},
    keywordstyle=\color{magenta},
    numberstyle=\tiny\color{codegray},
    stringstyle=\color{codepurple},
    basicstyle=\ttfamily\footnotesize,
    breakatwhitespace=false,         
    breaklines=true,                 
    captionpos=b,                    
    keepspaces=true,                 
    numbers=left,                    
    numbersep=5pt,                  
    showspaces=false,                
    showstringspaces=false,
    showtabs=false,                  
    tabsize=2
}

\title{Multireference equation-of-motion driven similarity renormalization group: theoretical foundations and applications to ionized states}

\author{Zijun Zhao}
\email{zijun.zhao@emory.edu}
\author{Shuhang Li}
\email{shuhang.li@emory.edu}
\author{Francesco A. Evangelista}
\email{fevange@emory.edu}
\affiliation{Department of Chemistry and Cherry Emerson Center for Scientific Computation, Emory University, Atlanta, Georgia, 30322, United States}

\date{\today}

\begin{document}
\begin{tocentry}
\includegraphics[width=3.25in]{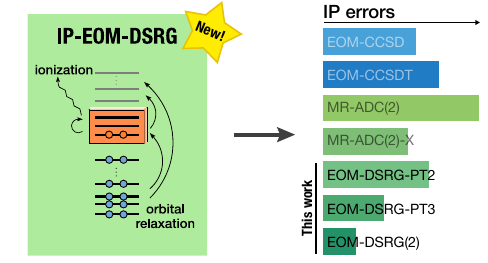}
\end{tocentry}

\begin{abstract}
	We present a formulation and implementation of an equation-of-motion (EOM) extension of the multireference driven similarity renormalization group (MR-DSRG) formalism for ionization potentials (IP-EOM-DSRG).
	The IP-EOM-DSRG formalism results in a Hermitian generalized eigenvalue problem, delivering accurate ionization potentials for systems with strongly correlated ground and excited states.
	The EOM step scales as $\mathcal{O}(N^5)$ with the basis set size $N$, allowing for efficient calculation of spectroscopic properties, such as transition energies and intensities.
	The IP-EOM-DSRG formalism is combined with three truncation schemes of the parent MR-DSRG theory: an iterative nonperturbative method with up to two-body excitations [MR-LDSRG(2)] and second- and third-order perturbative  approximations [DSRG-MRPT2/3].
	We benchmark these variants by computing 1) the vertical valence ionization potentials of a series of small molecules at both equilibrium and stretched geometries; 2) the spectroscopic constants of several low-lying electronic states of the \ch{OH}, \ch{CN}, \ch{N2+}, and \ch{CO+} radicals; and 3) the binding curves of low-lying electronic states of the \ch{CN} radical.
	A comparison with experimental data and theoretical results shows that all three IP-EOM-DSRG methods accurately reproduce the vertical ionization potentials and spectroscopic constants of these systems. Notably, the DSRG-MRPT3 and MR-LDSRG(2) versions outperform several state-of-the-art multireference methods of comparable or higher cost.
\end{abstract}

\maketitle
\section{Introduction}
The accurate and simultaneous description of multiple electronically excited states, especially in the presence of strong electron correlation, is an open challenge in quantum chemistry.
Existing strategies for calculating excited states lie between two philosophies: state-specific and equation-of-motion (EOM)-like methods.\cite{rowe.1968.10.1103/RevModPhys.40.153}
State-specific schemes optimize each target state separately, whereas EOM-like methods directly compute the energy differences between a reference state and the states of interest by solving a (generalized) eigenvalue problem in a basis of excitation operators.

State-specific methods include formalisms such as coupled-cluster (CC) theory,\cite{kutzelnigg.1982.10.1063/1.444231,crawford.2007.10.1002/9780470125915.ch2,shavitt.2009.10.1017/CBO9780511596834,shen.2012.10.1016/j.chemphys.2011.11.033,lee.2019.10.1063/1.5128795,tuckman.2023.10.1021/acs.jctc.3c00194,damour.2024.10.1021/acs.jctc.4c00034} and EOM-like methods include the well-known EOM-CC methods,\cite{krylov.2008.10.1146/annurev.physchem.59.032607.093602,bartlett.2012.10.1002/wcms.76,lischka.2018.10.1021/acs.chemrev.8b00244,musial.2020.10.1002/9781119417774.ch4} and also those based on the linear response (LR)\cite{sneskov.2012.10.1002/wcms.99}, propagator\cite{lowdin.1985.10.1016/S0065-32760860305-6,dreuw.2015.10.1002/wcms.1206,lefrancois.2015.10.1063/1.4931653}, and SAC-CI\cite{hirao.1978.10.1063/1.436452,nakatsuji.1981.10.1063/1.442386,nakatsuji.1997.10.1142/9789812812148_0002} formalisms, as they all result in similar working equations.
Configuration interaction (CI)\cite{knowles.1984.10.1016/0009-26148485513-X,sherrill.1999.10.1016/S0065-32760860532-8}  and its multireference generalization (MRCI)\cite{siegbahn.1980.10.1002/qua.560180510,werner.1988.10.1063/1.455556,knowles.1992.10.1007/BF01117405} may also be viewed as belonging to the EOM category, as they can target multiple states at once.

Intermediate between these two philosophies are methods based on the effective Hamiltonian theory.\cite{bloch.1958.10.1016/0029-55825890116-0,lindgren.1986.10.1007/978-3-642-61640-2} Most adopt the ``transform-then-diagonalize'' strategy, where reduced-dimension effective Hamiltonians containing interactions between the states of interest are formed \textit{via} certain transformations and diagonalized to obtain the desired states.
These encompass formalisms variously known as quasi-degenerate (QD),\cite{angeli.2004.10.1063/1.1778711,sharma.2016.10.1063/1.4939752} state-averaged (SA),\cite{li.2018.10.1063/1.5019793} multi-state (MS),\cite{finley.1998.10.1016/S0009-26149800252-8,shiozaki.2011.10.1063/1.3633329,li.2019.10.1063/1.5088120} or density-averaged\cite{aoto.2016.10.1063/1.4941604} methods.

Both state-specific and EOM-like approaches can be formulated as single-reference or multireference methods, depending on whether the reference state is a single Slater determinant or a correlated wavefunction.
The latter is necessary for describing strong electron correlation.\cite{lyakh.2012.10.1021/cr2001417}
Effective Hamiltonian methods, in contrast, are typically multireference schemes and are attractive options when a small number of states are of interest; however, they lack the flexibility of EOM-like methods to describe transitions involving orbitals other than frontier/active orbitals.

In the scenario where a large number of states are required, state-specific methods quickly become cost-prohibitive.
State properties, such as excitation energies, obtained from effective Hamiltonian theories are typically dependent on the number of states considered, and their accuracy tends to deteriorate when a large number of states are required.
EOM-like theories, on the other hand, do not suffer from this issue.
Therefore, EOM-like formalisms are preferred not only for their flexiblity in simulation of a wide range of spectroscopic processes starting from a single state, but also for their ability to describe a large number of states at a relatively modest computational cost.

The EOM formalism as applied to the coupled-cluster ansatz (EOM-CC) has enjoyed great popularity as it has given rise to a family of flexible, low-scaling, and systematically improvable methods that are essentially black-box.\cite{stanton.1993.10.1063/1.464746,nooijen.1997.10.1063/1.474000,levchenko.2004.10.1063/1.1630018,gour.2005.10.1063/1.2042452,bauman.2017.10.1080/00268976.2017.1350291}
In contrast, multireference (MR) extensions of EOM theory have progressed more slowly.
A fairly mature formalism is the multireference EOM-CC (MR-EOM-CC) methodology of Nooijen and coworkers,\cite{datta.2012.10.1063/1.4766361,demel.2013.10.1063/1.4796523,nooijen.2014.10.1063/1.4866795,huntington.2015.10.1063/1.4921187,huntington.2016.10.1021/acs.jctc.5b00799} whose ground state is a partially internally contracted multireference coupled-cluster (pIC-MRCC) state,\cite{datta.2011.10.1063/1.3592494} and the excited states are found by diagonalizing the similarity-transformed Hamiltonian in a compact uncontracted basis (of determinants).
However, the MR-EOM-CC methodology has more in common with effective Hamiltonian theories, such as density-averaged multistate internally contracted MRCC theory,\cite{aoto.2016.10.1063/1.4941604} and the state-averaged SA-DSRG formalism\cite{li.2018.10.1063/1.5019793}, than with EOM formalisms.
Furthermore, this method has so far been applied exclusively to neutral excitations, and no IP extension has been reported.
Linear-response variants of projective MRCC method, namely Mukherjee multireference coupled cluster (Mk-MRCC-LR) \cite{jagau.2012.10.1063/1.4734308,jagau.2012.10.1063/1.4734309} and internally-contracted multireference coupled cluster (ic-MRCC-LR),\cite{samanta.2014.10.1063/1.4869719} can be more properly categorized as MR EOM-like methods.

Another family of MR EOM-like methods belong to the multireference algebraic diagrammatic construction (MR-ADC) formalism by Sokolov and coworkers.\cite{sokolov.2018.10.1063/1.5055380,chatterjee.2019.10.1021/acs.jctc.9b00528,chatterjee.2020.10.1021/acs.jctc.0c00778,demoura.2024.10.1021/acs.jpca.4c03161,gaba.2024.10.1039/D4CP00801D,mazin.2023.10.1021/acs.jctc.3c00477,moura.2022.10.1039/D1CP05476G,mazin.2021.10.1021/acs.jctc.1c00684} 
In MR-ADC, the ground state is obtained from partially internally contracted second-order $N$-electron valence state perturbation theory (pc-NEVPT2),\cite{angeli.2001.10.1016/S0009-26140101303-3,angeli.2001.10.1063/1.1361246,angeli.2004.10.1063/1.1778711,sokolov.2024.10.48550/arXiv.2401.11262} also known as fully internally contracted (FIC-) NEVPT2.\cite{guo.2021.10.1063/5.0051211}
The excited states are obtained by diagonalizing the Hermitian effective Liouvillian in a basis of internally contracted configurations.
The effective Liouvillian formalism\cite{mukherjee.1989.10.1007/978-3-642-93424-7_12} used in MR-ADC is distinct from but related to the EOM approach.

The authors of this work have recently reported an MR EOM method based on the ic-MR unitary CC approach (EOM-ic-MRUCC),\cite{li.2025.10.1063/5.0261000} which provides the theoretical foundation for the present work.
Here, we present the EOM extension of the MR driven similarity renormalization group (MR-DSRG) formalism.
The MR-DSRG formalism can be viewed as a renormalized version of the ic-MRUCC theory,\cite{hoffmann.1987.10.1016/0009-26148780642-5,hoffmann.1988.10.1063/1.454125,chen.2012.10.1063/1.4731634} which additionally avoids issues surrounding the orthogonalization of the internally contracted excited configurations by using a set of so-called \textit{many-body conditions} to determine the ground state wavefunction (see \cref{sec:mrdsrg} for details).
We formulate a simple MR generalization of the EOM formalism for computing ionization potentials for three methods based on MR-DSRG theory, including the perturbative DSRG-MRPT2/3 methods,\cite{li.2017.10.1063/1.4979016,li.2018.10.1063/1.5019793} and the iterative MR-LDSRG(2) method.\cite{li.2016.10.1063/1.4947218}
In \cref{sec:theory}, we briefly recapitulate the MR-DSRG and EOM formalisms, and discuss the derivation of the IP-EOM-DSRG formalism.
In \cref{sec:comp}, we present the implementation of the IP-EOM-DSRG methods, and discuss the computational details of the calculations.
Benchmark results appear in \cref{sec:results}, followed by conclusions and future directions in \cref{sec:conclusions}.

\section{Theory}
\label{sec:theory}
\subsection{The multireference DSRG formalism}
\label{sec:mrdsrg}
The MR-DSRG is a numerically robust internally contracted multireference electron correlation formalism.\cite{li.2019.10.1146/annurev-physchem-042018-052416}
Detailed accounts of the MR-DSRG formalism and associated methods can be found elsewhere,\cite{li.2016.10.1063/1.4947218,li.2019.10.1146/annurev-physchem-042018-052416} and we will only briefly recapitulate its salient features here.
The reference state in the MR-DSRG formalism is a complete active space (CAS) wavefunction, given by
\begin{equation}
	\ket{\Phi_0}=\sum_{\mu=1}^dc_{\mu}\ket{\phi_{\mu}}.
\end{equation}
The set of determinants $\msp = \{\ket{\phi_{\mu}}\},\mu=1,\dots,d$ defines the model space.
We assume that the orbitals have been partitioned into core ($\mathbf{C}$, indexed by $m,n,\dots$), active ($\mathbf{A}$, indexed by $u,v,\dots$), and virtual ($\mathbf{V}$, indexed by $e,f,\dots$) subsets.
We further introduce the hole ($\mathbf{H}=\mathbf{C}\cup\mathbf{A}$, indexed by $i,j,\dots$), and particle ($\mathbf{P}=\mathbf{A}\cup\mathbf{V}$, indexed by $a,b,\dots$) composite orbital spaces.
The MR-DSRG is formulated around a continuous unitary transformation, controlled by a time-like quantity, $s$, called the \textit{flow parameter}, described by
\begin{equation}
\label{eq:dsrgflow}
    \hat{H}\rightarrow\bar{H}(s)=e^{-\hat{A}(s)}\hat{H}e^{\hat{A}(s)},
\end{equation}
where $\bar{H}(s)$ is the MR-DSRG similarity-transformed Hamiltonian, and the anti-Hermitian operator $\hat{A}(s)$ is expressed in terms of a cluster operator as $\hat{A}(s)=\hat{T}(s)-\hat{T}(s)^{\dagger}$.
$\hat{T}(s)$ is parameterized as in the internally contracted generalization of coupled cluster theory,\cite{datta.2011.10.1063/1.3592494,evangelista.2011.10.1063/1.3559149,evangelista.2012.10.1063/1.4718704,hanauer.2011.10.1063/1.3592786} with $s$-dependent amplitudes.
The flow parameter $s$ is henceforth omitted for brevity.

This transformation aims at making the Hamiltonian increasingly block-diagonal with a judicious choice of $\hat{A}$, by gradually suppressing the couplings between the reference wavefunction $\ket{\Phi_0}$ and the internally contracted excited configurations $\{\hat{a}^{ab\dots}_{ij\dots}\}\ket{\Phi_0} \equiv \{\hat{a}_{a}^\dagger \hat{a}_{b}^\dagger \dots \hat{a}_{j} \hat{a}_{i} \ket{\Phi_0}$, where the curly braces indicate generalized normal ordering,\cite{kutzelnigg.1997.10.1063/1.474405,mukherjee.1997.10.1016/S0009-26149700714-8} used throughout this work.
Since the hole-particle non-diagonal part of $\barh$ contains all operators responsible for these couplings, we partition $\barh$ into the sum of the diagonal part, $\barh^{\mathrm{D}}$, and the non-diagonal part, $\barh^{\mathrm{N}}$.
Instead of determining the $\hat{A}$ in \cref{eq:dsrgflow} projectively, as is done in most (MR)CC methods, it is determined by a set of \textit{many-body conditions} as follows:
\begin{equation}
	\label{eq:manybody-cond}
	\barh^{\mathrm{N}} = [e^{-\hat{A}}\hat{H}e^{\hat{A}}]^{\mathrm{N}} = \hat{R},
\end{equation}
where $\hat{R}$ is a source operator that drives the limit process such that $\lim_{s\rightarrow\infty}\hat{R}(s)=0$.
In the MR-DSRG formalism, an analytical expression of the source operator is derived by a second-order perturbative analysis of the single-reference SRG,\cite{evangelista.2014.10.1063/1.4890660} making \cref{eq:manybody-cond} a non-linear operator equation for the amplitudes of the cluster operator.

Practical realizations of the MR-DSRG method use either iterative or perturbative approximations.
In iterative methods, the cluster operator $\hat{T}$ is truncated to a given excitation rank, and the nonterminating Baker--Campbell--Hausdorff (BCH) expansion of the similarity-transformed Hamiltonian is approximated with a finite number of terms.
An iterative method based on the MR-DSRG formalism is MR-LDSRG(2),\cite{li.2016.10.1063/1.4947218} where the cluster operator is truncated to double excitations, and the similarity-transformed Hamiltonian is calculated with the linearized commutator approximation,\cite{yanai.2006.10.1063/1.2196410} where each commutator entering into the BCH expansion is truncated after two-body components as follows:
\begin{equation}
	\barh_{1,2}=\hat{H}+\sum_{k=1}^{\infty}\frac{1}{k!}\underbrace{[\dots[[\hat{H},\hat{A}]_{1,2},\hat{A}]_{1,2}\dots]_{0,1,2}}_{k\text{ nested commutators}},
\end{equation}
where $O_{1,2}$ indicates that up to two-body components are retained for $O$.
Alternatively, a perturbative analysis of the transformed Hamiltonian can be used to obtain one-shot, perturbative methods.\cite{li.2019.10.1146/annurev-physchem-042018-052416}
Second-\cite{li.2015.10.1021/acs.jctc.5b00134} and third-order\cite{li.2017.10.1063/1.4979016} perturbative MR-DSRG methods (DSRG-MRPT2/3) have also been developed, which produce similarity-transformed Hamiltonians that are correct up to second- and third-order in perturbation theory, respectively.

\subsection{EOM-DSRG formalism}
\label{sec:eomdsrg}
The EOM formalism was first proposed by Rowe in the context of nuclear spectroscopy.\cite{rowe.1968.10.1103/RevModPhys.40.153}
A slew of developments in the late 1970s to 1980s mainly focused on the linear response and symmetry-adapted cluster CI formalisms, but yielded working equations resembling those of the EOM working equations.\cite{monkhorst.1977.10.1002/qua.560120850,hirao.1978.10.1063/1.436452,paldus.1978.10.1103/PhysRevA.17.805,nakatsuji.1978.10.1016/0009-26147889113-1,nakatsuji.1978.10.1063/1.436028,nakatsuji.1981.10.1063/1.442386,ghosh.1981.10.1080/00268978100101261,emrich.1981.10.1016/0375-94748190179-2,emrich.1981.10.1016/0375-94748190180-9,ghosh.1982.10.1016/0301-01048287077-8,dalgaard.1983.10.1103/PhysRevA.28.1217,sekino.1984.10.1002/qua.560260826} 
The EOM-CC theory as we know it today took shape from the late 1980s onwards.\cite{geertsen.1989.10.1016/0009-26148985202-9, stanton.1993.10.1063/1.464746,comeau.1993.10.1016/0009-26149389023-B, nooijen.1995.10.1063/1.469147, watts.1996.10.1016/0009-26149600708-7}
Following the classic EOM formalism, we define the EOM-DSRG ansatz as follows:
\begin{equation}
	\ket{\Psi_{\alpha}} = \bar{\mathcal{R}}_{\alpha}\ket{\Psi_0},
\end{equation}
where $\bar{\mathcal{R}}_{\alpha}$ is a \textit{state-transfer} operator delivering the $\alpha$-th excited state ($\Psi_{\alpha}$) from the (in principle exact) ground state, \textit{i.e.}, formally $\bar{\mathcal{R}}_{\alpha}\equiv\ket{\Psi_{\alpha}}\bra{\Psi_0}$.
To realize a variational optimization of $N$ electronic states, we introduce an energy functional augmented with orthonormality constraints:
\begin{equation}
	\label{eq:eom-variational}
\mathcal{L} = \sum_\alpha^{N} \braket{\Psi_0|\bar{\mathcal{R}}_{\alpha}\adj\hat{H}\bar{\mathcal{R}}_{\alpha}|\Psi_0} - \sum_{\alpha\beta}^{N} \lambda_{\alpha\beta} (\braket{\Psi_0|\bar{\mathcal{R}}\adj_{\beta}\bar{\mathcal{R}}_{\alpha}|\Psi_0} - \delta_{\alpha \beta}).
\end{equation}
Here, we assume that all excited states are orthogonal to the ground state ($\braket{\Psi_0|\Psi_\alpha} = \braket{\Psi_0|\bar{\mathcal{R}}_{\alpha}|\Psi_0} = 0$), which we achieve by careful parameterization of $\bar{\mathcal{R}}_{\alpha}$.

Due to its formal advantages, we adopt the self-consistent excitation operators introduced by Mukherjee and coworkers, \cite{prasad.1985.10.1103/PhysRevA.31.1287, datta.1993.10.1103/PhysRevA.47.3632} which expresses $\bar{\mathcal{R}}_{\alpha}$ as a similarity-transformed operator:
\begin{equation}
    \bar{\mathcal{R}}_{\alpha}\equiv e^{\hat{A}}\hat{\mathcal{R}}_{\alpha}e^{-\hat{A}}.
\end{equation}
Substituting this expression into \cref{eq:eom-variational}, we arrive at a simpler and equivalent energy functional that involves expectation values with respect to the reference state and the similarity-transformed Hamiltonian ($\bar{H} = e^{-\hat{A}} \hat{H} e^{\hat{A}}$):
\begin{equation}
\label{eq:eom-master}
\mathcal{L} = \sum_\alpha^{N} \braket{\Phi_0|\hat{\mathcal{R}}_{\alpha}\adj\bar{H}\hat{\mathcal{R}}_{\alpha}|\Phi_0} - \sum_{\alpha\beta}^{N} \lambda_{\alpha\beta} (\braket{\Phi_0|\hat{\mathcal{R}}\adj_{\beta}\hat{\mathcal{R}}_{\alpha}|\Phi_0} - \delta_{\alpha \beta}).
\end{equation}

By convention, $\hat{\mathcal{R}}_{\alpha}$ is parameterized as a linear excitation operator (\textit{i.e.}, a linear combination of excitation operators, as opposed to a non-linear, \textit{e.g.}, exponential, parameterization).\cite{harris.1977.10.1002/qua.560120848,sekino.1984.10.1002/qua.560260826}
Therefore we can write $\hat{\mathcal{R}}_{\alpha}$ as a linear combination of excitation operators $\{\hat{{\rho}}_p\}$ with corresponding excitation amplitudes $r^p_{\alpha}$:
\begin{equation}
	\hat{\mathcal{R}}_{\alpha} = \sum_{p = 1}^{n_{\mathrm{eom}}}r_{\alpha}^{p}\hat{\rho}_p,
\end{equation}
where $n_{\mathrm{eom}}$ is the number of excitation operators.

Imposing stationarity of the energy functional \cref{eq:eom-master} with respect to variations of the excitation amplitudes $r_{\alpha}^p$ (taken to be real) and the Lagrange multipliers $\lambda_{\alpha\beta}$, following a derivation similar to that of the canonical Hartree--Fock equations,\cite{szabo.1996.} one can show the following generalized eigenvalue problem arises:
\begin{equation}
	\sum_{q=1}^{n_{\mathrm{eom}}}\braket{\Phi_0|\hat{\rho}_p^{\dagger}\bar{H}\hat{\rho}_q|\Phi_0} r_{\alpha}^q
	= E_{\alpha} \sum_{q=1}^{n_{\mathrm{eom}}} \braket{\Phi_0|\hat{\rho}_p^{\dagger}\hat{\rho}_q|\Phi_0} r_{\alpha}^q,
\label{eq:eom_tht}
\end{equation}
where $E_{\alpha}$ is the excited state energy, from which the excitation energies $\omega_\alpha = E_\alpha - E_0$, $E_0=\braket{\Phi_0|\barh|\Phi_0}$ can be obtained.

Clearly, if $\barh$ is Hermitian, then \cref{eq:eom_tht} is a Hermitian generalized eigenvalue problem.
The orthogonality between each excited state and the ground state is enforced by the following conditions:
\begin{equation}
	\braket{\Phi_0|\hat{\mathcal{R}}_{\alpha}|\Phi_0} = 0.
\end{equation}
These conditions are satisfied by our choice of excitation operators, as we will show in \cref{sec:IP-EOM-dsrg}, while the orthonormality constraint is automatically satisfied by solutions of the generalized eigenproblem in \cref{eq:eom_tht}.

The EOM-DSRG formalism is agnostic to the particular choice of underlying MR-DSRG method, and we investigate the performance of all three choices of MR-DSRG methods in this work. 
We will refer to the EOM-DSRG method based on the MR-LDSRG(2) simply as EOM-DSRG(2), and those based on DSRG-MRPT2/3 as EOM-DSRG-PT2/3.
The MR-DSRG formalism also allows for state-specific and state-averaged calculations of electronically excited states,\cite{li.2018.10.1063/1.5019793} so the EOM-DSRG formalism can simulate more sophisticated spectroscopic processes that go through electronically excited states, such as resonance-enhanced multiphoton ionization (REMPI).

Before closing this section, we would like to offer some comments on the specific choices made in deriving the EOM-DSRG formalism.
In many conventional EOM theories, the excitation energy $\omega_{\alpha}$ is computed by taking difference between the Schr{\"o}dinger equation for the ground and excited states, leading to the condition:
\begin{equation}
	[\bar{H},\hat{\mathcal{R}}_{\alpha}]\ket{\Phi_0}=\omega_{\alpha}\hat{\mathcal{R}}_{\alpha}\ket{\Phi_0}.
\end{equation}
The corresponding equations for the excitation amplitudes are a generalized eigenvalue problem:
\begin{equation}
\label{eq:eom_single_comm}
	\sum_{q=1}^{n_{\mathrm{eom}}}\braket{\Phi_0|\hat{\rho}_p^{\dagger}[\bar{H},\hat{\rho}_q]|\Phi_0} r_{\alpha}^q
	= \omega_{\alpha} \sum_{q=1}^{n_{\mathrm{eom}}} \braket{\Phi_0|\hat{\rho}_p^{\dagger}\hat{\rho}_q|\Phi_0} r_{\alpha}^q,
\end{equation}
which we call the single-commutator form of the EOM equations.
The equivalence of \cref{eq:eom_tht} and \cref{eq:eom_single_comm} is guaranteed only when the ground state is determined projectively (like in ic-MRUCC\cite{li.2025.10.1063/5.0261000}) but not in truncated MR-DSRG approaches.
To see this, we make use of a resolution of identity in terms of the projector, $\hat{P}=\sum_{\mu}\ket{\Phi_{\mu}}\bra{\Phi_{\mu}}$, onto the model space $\mathcal{M}$, and the projector onto its orthogonal complement, $\hat{Q}=1-\hat{P}$.
The additional term introduced by the commutator in the LHS of \cref{eq:eom_single_comm}, $\braket{\Phi_0|\hat{\rho}_p^{\dagger}\hat{\rho}_q\bar{H}|\Phi_0}$, can be expressed as
\begin{equation}
	\braket{\Phi_0|\hat{\rho}_p^{\dagger}\hat{\rho}_q\barh|\Phi_0} = \braket{\Phi_0|\hat{\rho}_p^{\dagger}\hat{\rho}_q(\hat{P}\barh\hat{P}+\hat{Q}\barh\hat{P})|\Phi_0}.
\end{equation}
Assuming $\ket{\Phi_0}$ is an eigenfunction of $\hat{P}\barh\hat{P}$, the first term reduces to $E_0\braket{\Phi_0|\hat{\rho}_p^{\dagger}\hat{\rho}_q|\Phi_0}$, where $E_0$ is the eigenvalue.
The second term will vanish if the ground state is determined projectively, resulting in zero coupling between the model space and its orthogonal complement\bibnote{This is true assuming no truncation of the cluster operator. In single-reference EOM-CC, this can still hold true for truncated cluster and EOM operators, since determinants of different excitation ranks are orthogonal, which means for EOM-CCSD, $\bra{\Phi_0}\hat{\rho}_p\adj\hat{\rho}_q\hat{Q}_{\mathrm{TQ\dots}}=0$, where $\hat{Q}_{\mathrm{TQ\dots}}$ is the projector onto triple and higher excitations. In the MR case, decoupling of the model space and its orthogonal complement occurs if the lower excitations have been projected out of the higher excitations, which, in practice, is not done for excitation ranks higher than the truncation rank.}, \textit{i.e.}, $\hat{Q}\barh\hat{P}\ket{\Phi_0}=0$.
This enables us to simplify \cref{eq:eom_single_comm} to
\begin{equation}
	\sum_{q=1}^{n_{\mathrm{eom}}}\braket{\Phi_0|\hat{\rho}_p^{\dagger}\bar{H}\hat{\rho}_q|\Phi_0} r_{\alpha}^q
	= E_{\alpha} \sum_{q=1}^{n_{\mathrm{eom}}} \braket{\Phi_0|\hat{\rho}_p^{\dagger}\hat{\rho}_q|\Phi_0} r_{\alpha}^q,
\end{equation}
which is equivalent to \cref{eq:eom_tht}.
In the MR-DSRG theory, the ground state is determined by the many-body conditions [\cref{eq:manybody-cond}], which results in a non-vanishing $\hat{Q}\barh\hat{P}\ket{\Phi_0}$, and in turn, makes the single-commutator form not equivalent to \cref{eq:eom_tht}.
For theories that do not satisfy the projective condition, the single-commutator form of the EOM equations will not be Hermitian, even when $\barh$ is Hermitian.
One may still formulate a Hermitian eigenvalue problem by symmetrizing the single commutator form of the EOM equations, which can be justified when the residuals $\hat{Q}\barh\hat{P}\ket{\Phi_0}$ are small, \textit{i.e.}, in the case of variational UCC.\cite{kim.2023.10.1021/acs.jpca.3c02480}
However, we do not consider this approach in this work, as the MR-DSRG residuals are not guaranteed to be small.

In many works, one may also find the double-commutator form of EOM equations given by
\begin{equation}
	\label{eq:eom_double_comm}
		\sum_{q=1}^{n_{\mathrm{eom}}}\braket{\Phi_0|[\hat{\rho}_p^{\dagger},[\bar{H},\hat{\rho}_q]]|\Phi_0} r_{\alpha}^q
		= \omega_{\alpha} \sum_{q=1}^{n_{\mathrm{eom}}} \braket{\Phi_0|[\hat{\rho}_p^{\dagger},\hat{\rho}_q]|\Phi_0} r_{\alpha}^q.
	\end{equation}
This may be preferred over the single-commutator form because it is manifestly connected and leads to lower-rank reduced density cumulants in the multireference case.
The equivalence of the single- and double-commutator formulations hinges on satisfying the so-called killer condition, which requires $\hat{\rho}_p^{\dagger}\ket{\Phi_0} = 0$, $\forall p$.
Since the EOM-DSRG formalism does not make use of either the single- or double-commutator form, we do not require the killer condition to be satisfied.
Interested readers are referred to the literature for detailed discussions.\cite{goscinski.1980.10.1088/0031-8949/21/3-4/026, weiner.1980.10.1002/qua.560180417, prasad.1985.10.1103/PhysRevA.31.1287, datta.1993.10.1103/PhysRevA.47.3632, szekeres.2001.10.1039/B008428J}
In the closely related EOM-ic-MRUCC formalism (Ref.~\citenum{li.2025.10.1063/5.0261000}), the ground state is obtained projectively, and we ensured that all operators in the excitation operator satisfy the killer condition \textit{via} a small modification to the internal excitation operators, so that we could adopt the double-commutator form of the EOM equations.
Finally, the double-commutator form of the EOM equations is also not Hermitian for theories not satisfying the projective condition, and it may be acceptable to symmetrize the theory when the projective residuals are small.\cite{kim.2023.10.1021/acs.jpca.3c02480}

Given the benefits of the double-commutator form, it might seem unreasonable for us to use \cref{eq:eom_tht} as our working equation for the EOM-DSRG formalism.
However, we have several reasons for this choice.
Firstly, the diagrams are already naturally connected in EOM-DSRG, since all many-body operators are normal ordered.
Secondly, rank reduction can be introduced on an \textit{ad hoc} basis without approximation, as will be discussed in \cref{sec:IP-EOM-dsrg}.
Lastly, as discussed above, the double-comutator Hamiltonian matrix would not be Hermitian in EOM-DSRG, and \textit{ad hoc} Hermitization cannot be justified.

\subsection{IP-EOM-DSRG}
\label{sec:IP-EOM-dsrg}
We now discuss the form of the excitation operator $\hat{\mathcal{R}}_{\alpha}$ used to compute singly ionized states in the IP-EOM-DSRG scheme.
We partition the EOM operator into an internal ($\hat{\mathcal{R}}^{\mathrm{int}}_{\alpha}$) and external ($\hat{\mathcal{R}}^{\mathrm{ext}}_{\alpha}$) part:
\begin{equation}
\hat{\mathcal{R}}_{\alpha} = 
\hat{\mathcal{R}}^{\mathrm{int}}_{\alpha}
+ \hat{\mathcal{R}}^{\mathrm{ext}}_{\alpha} ,
\end{equation}
where the internal part maps the parent model space into the target model space ($\hat{\mathcal{R}}^{\mathrm{int}}_{\alpha} \msp \in \msp^\mathrm{1h}$), where $\msp^\mathrm{1h}$ is the model space formed by removing one active electron from the parent model space determinants in all possible ways.
The external part generates excited configurations outside the target model space ($\hat{\mathcal{R}}^{\mathrm{ext}}_{\alpha} \msp \notin \msp^\mathrm{1h}$). 
Operators from these two groups span orthogonal spaces when applied to the state $\ket{\Phi_0}$.
No scalar term enters in $\hat{\mathcal{R}}_{\alpha}$ as it mixes in a different sector of Fock space, also guaranteeing orthogonality with the ground state.

We parameterize external excitations with the set of one-hole (1h) and two-hole-one-particle (2h1p) operators, excluding those labeled only by active indices (denoted by $'$):
\begin{equation}
\hat{\mathcal{R}}^{\mathrm{ext}} = \sum_{i}^{\mathbf{H}}{}^{'}r^i\{\hat{a}_i\}+\frac{1}{2}\sum_{ij}^{\mathbf{H}}{}^{'}\sum_{a}^{\mathbf{P}}{}^{'}r^{ij}_a\{\hat{a}^{a}_{ij}\}
\end{equation}
where the curly braces indicate generalized normal ordering, and the state index $\alpha$ is omitted for clarity.

We use the many-body (MB) form of internal excitations with up to 2h1p operators:
\begin{align}
\hat{\mathcal{R}}^{\mathrm{int}}_\text{MB} =& \sum_{ux}^{\textbf{A}}r_{}^{u} \{ \hat{a}^{}_{u} \} + \frac{1}{2}\sum_{uvx}^{\textbf{A}}r_{x}^{uv}\{ \hat{a}^{x}_{uv} \}.
\end{align}
This choice ensures that the number of internal excitation operators scales polynomially with the number of active orbitals, and also retains invariance to unitary transformations within the active space.
Other choices of internal excitation operators are possible, such as the active-space eigenoperators adopted by Sokolov and coworkers for the MR-ADC formalism.\cite{sokolov.2018.10.1063/1.5055380}
A drawback of the eigenoperators is that they formally scale exponentially with the number of active orbitals, and therefore truncation is necessary for a practical implementation; additionally, transition RDMs between the ground and excited states need to be evaluated, leading to higher storage requirements.
However, the eigenoperators can describe higher-order internal excitations than MB operators in cases where the MB operators cannot saturate the target model space.
We return to this point in \cref{sec:vert_ip}.

\begin{figure}[!htb]
	\centering
	\includegraphics[width=6.5in]{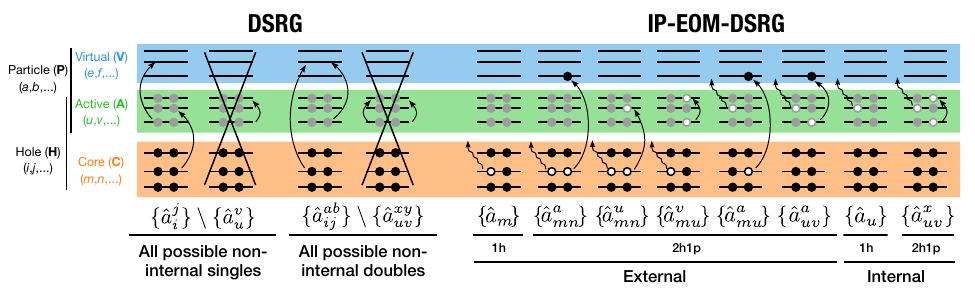}
	\caption{Definitions of the cluster operator ($\hat{T}$, left) and the EOM excitation operator ($\hat{\mathcal{R}}$, right) used in the multireference formulation of the DSRG and IP-EOM-DSRG schemes. Curved arrows indicate the excitation of an electron, while wavy arrows indicate the ionization of an electron.}
	\label{fig:operators}
\end{figure}

In all, the IP-EOM-DSRG excitation operator is given by:
\begin{equation}
	\hat{\mathcal{R}} = \sum_{i}^{\mathbf{H}}r^i\{\hat{a}_i\}+\frac{1}{2}\sum_{ij}^{\mathbf{H}}\sum_{a}^{\mathbf{P}}r^{ij}_a\{\hat{a}^{a}_{ij}\},
\end{equation}
which is a transparent generalization of the single-reference IP-EOM-CCSD ansatz.\cite{nooijen.1995.10.1063/1.468900, stanton.1995.10.1063/1.469817}
The use of generalized normal ordering automatically ensures the orthogonality between ground and excited states (see \cref{sec:eomdsrg}).
As a side note, MR-DSRG theory is formulated in a semicanonical orbital basis, which is not uniquely defined when there is orbital degeneracy, but the MR-DSRG formalism is invariant to unitary transformations within sets of degenerate orbitals.\cite{li.2019.10.1146/annurev-physchem-042018-052416}
Our choice of the IP-EOM-DSRG excitation operator basis preserves this invariance.

We can write the matrix representation of the IP-EOM-DSRG equation as:
\begin{equation}
	\bar{\mathbf{H}} \mathbf{r}_{\alpha} = \omega_{\alpha} \mathbf{S}\mathbf{r}_{\alpha}, \label{eq:matrix eom}
\end{equation} 
where the transformed Hamiltonian matrix is given by:
\begin{equation}
\bar{H}_{pq} =\braket{\Phi_0|\hat{\rho}_p^{\dagger}\bar{H}\hat{\rho}_q|\Phi_0},	
\end{equation}
and the metric matrix is: 
\begin{equation}
S_{pq} =\braket{\Phi_0|\hat{\rho}_p^{\dagger}\hat{\rho}_q|\Phi_0}.
\end{equation}
The set of internally contracted excited configurations $\{\hat{\rho}_p\ket{\Phi_0}\}$ exhibits linear dependence and can be orthogonalized via a linear transformation induced by the matrix $\mathbf{M}$,
\begin{equation}
	\boldsymbol{\hat{\chi}} = \boldsymbol{\hat{\rho}}\mathbf{M}, \label{eq:generalized eigen start}
\end{equation}
where $\boldsymbol{\hat{\chi}}$ is the row-vector of orthogonalized operators, and $\boldsymbol{\hat{\rho}}$ is the row-vector of the original bare excitation operators.
The matrix $\mathbf{M}$ depends parametrically on some threshold(s) $\bvec{\eta}$ for the orthogonalization procedure, and interested readers are referred to detailed discussions in the literature.\cite{evangelista.2012.10.1063/1.4718704,hanauer.2012.10.1063/1.4757728,sokolov.2018.10.1063/1.5055380,kohn.2020.10.1080/00268976.2020.1743889,moura.2022.10.1039/D1CP05476G}
In this work, we use the two-threshold Gram--Schmidt sequential orthogonalization procedure proposed by K{\"o}hn and coworkers.\cite{kohn.2020.10.1080/00268976.2020.1743889}
When expressed in the $\boldsymbol{\hat{\chi}}$ operator basis, \cref{eq:matrix eom} becomes an ordinary eigenvalue problem:
\begin{equation}
	\bar{\mathbf{H}}' \mathbf{r}_{\alpha}' = \omega_\alpha \mathbf{r}_{\alpha}',\label{eq:ge3}
\end{equation}
where the modified matrix $\bar{\mathbf{H}}'$ and vector $\mathbf{r}'_{\alpha}$ are defined as	$\mathbf{\barh}' = \mathbf{M}^{\dagger}\bar{\mathbf{H}}\mathbf{M}$
and $\mathbf{r}_{\alpha}' = \mathbf{M}^{\dagger}\mathbf{r}_{\alpha}$.

A na{\"i}ve evaluation of the matrix elements $\barh_{pq}$ where $p,q\in\{\hat{\mathcal{R}}^{\mathrm{int}}_{\mathrm{MB}}\}$ can involve up to $5$-body reduced density cumulants, with prohibitive computational and storage costs. For example, the matrix element of $\bar{H}$ computed with respect to two internal 2h-1p states contains the leading term:
\begin{equation}
	\braket{\Phi_0|\{\hat{a}^{uv}_{w}\}\barh^{pq}_{rs}\{\hat{a}^{rs}_{pq}\}\{\hat{a}^{x}_{yz}\}|{\Phi_0}}=\barh^{pq}_{rs}\lambda^{uvrsx}_{wpqyz}+\dots.
\end{equation}
To avoid this, we can write
\begin{equation}
\label{eq:commutator_trick}
	\braket{\Phi_0|\hat{\rho}_p^{\dagger}\barh\hat{\rho}_q|\Phi_0} = \braket{\Phi_0|\hat{\rho}_p^{\dagger}[\barh,\hat{\rho}_q]|\Phi_0}+\braket{\Phi_0|\hat{\rho}_p^{\dagger}\hat{\rho}_q\barh|\Phi_0},
\end{equation}
where the introduction of the commutator in the first term allows us to utilize the rank-reducing property of the commutator, such that only up to $4$-body reduced density cumulants are needed.
This trick was first proposed by Dyall,\cite{dyall.1995.10.1063/1.469539} and is used in most variants of NEVPT2,\cite{angeli.2001.10.1016/S0009-26140101303-3,angeli.2005.10.1016/j.cpc.2005.05.002,guo.2021.10.1063/5.0051211} and ic-MRCI methods.\cite{saitow.2013.10.1063/1.4816627,sivalingam.2016.10.1063/1.4959029} 
If both $\hat{\rho}_p$ and $\hat{\rho}_q$ belong to the same class of operators (in this case, we are only interested in internal operators as they introduce the highest rank cumulants), then $\hat{\rho}_p\adj\hat{\rho}_q\ket{\Phi_0}\in \msp$.
If furthermore $\ket{\Phi_0}$ is an eigenstate of $\hat{P}\barh\hat{P}$, then the second term $\braket{\Phi_0|\hat{\rho}_p^{\dagger}\hat{\rho}_q\barh|\Phi_0} = E_0 S_{pq}$, which amount to a shift in the eigenvalues by $E_0$.
For this condition to be met, the MR-LDSRG(2) effective Hamiltonian must be diagonalized in the CAS space to obtain a reference state $\ket{\Phi_0}$ that is an eigenstate of $\hat{P}\barh\hat{P}$.
We then use the $\barh$ and the reduced density cumulants of $\ket{\Phi_0}$ in the subsequent IP-EOM-DSRG calculations.

Numerical experiments show that the rigorous evaluation of contributions containing four-body density cumulants is critical to avoid artificial variational collapse of the EOM energies, as has been reported before in cumulant-truncated ic-MRCI methods.\cite{saitow.2013.10.1063/1.4816627}
The resulting IP-EOM-DSRG equations depend on up to the $4$-body reduced density cumulants.
This procedure is used throughout this work, as it is essential for a computationally feasible IP-EOM-DSRG method, and involves no approximations.
Note that since the killer condition is not satisfied by the many-body parameterization of internal excitations, it is not possible to proceed one step further and convert terms like those in \cref{eq:commutator_trick} into a double commutator form.

Finally, we note that the IP-EOM-DSRG method will not deliver size-intensive excitation energies.
Size intensivity describes when excitation energies remain constant in the presence of non-interacting subsystems, and it was shown in Appendix B of Ref.~\citenum{li.2025.10.1063/5.0261000} that a necessary condition for size intensivity is that the ground state is determined projectively, which is not the case for the MR-DSRG methods.
In the case of IP-EOM-DSRG with up to 2h-1p excitations, only projections onto single excitations must be zero to guarantee size intensivity.
In practice, we find that the size-intensitivity errors are small for realistic systems, as we report in \cref{sec:size_intensivity}.

\subsection{Calculation of transition intensities}
The intensity of a photoelectron transition at energy $\omega$, $I(\omega)$, can be obtained by computing the photoionization cross section, $\sigma(\omega)$, which is proportional to the differential oscillator strength, $f(\omega)$ of a transition to a continuum state at energy $\omega$.\cite{cukras.2014.10.1063/1.4900545,gozem.2015.10.1021/acs.jpclett.5b01891}
This requires the explicit consideration of Dyson orbitals, which is beyond the scope of the current study.
Instead, we opt to use the spectroscopic factor, $S_{\alpha}$ (sometimes also denoted as $P_{\alpha}$), as a proxy for $I(\omega_{\alpha})$.\cite{chatterjee.2019.10.1021/acs.jctc.9b00528}
This quantity is given as follows:\cite{meldner.1971.10.1103/PhysRevA.4.1388,kheifets.1994.10.1088/0953-4075/27/15/005}
\begin{equation}
\label{eq:Sk}
	S_k=\sum_{i}^{\bvec{H}}|\braket{\Psi^{N-1}_{\alpha}|\hat{a}_i|\Psi_0^{N}}|^2.
\end{equation}
$S_k$ is defined as the sum of ionization probabilities from all orbitals of the parent ground state that result in the ${\alpha}$-th excited state of the ionized system, and is therefore related to the intensity of a photoelectron transition.
In \cref{eq:Sk}, $\ket{\Psi^{N-1}_{\alpha}}$ is the ${\alpha}$-th excited state of the ionized system, and $\ket{\Psi_0^{N}}$ is the ground state of the parent system.
Substituting in the IP-EOM-DSRG ansatz for the excited state, we have
\begin{align}
	S_{\alpha}&=\sum^{\bvec{H}}_i|\braket{\Psi_0^N|\bar{\mathcal{R}}_{\alpha}\adj\hat{a}_i|\Psi_0^N}|^2\\
	   &=\sum^{\bvec{H}}_i|\braket{\pz|\hat{\mathcal{R}}_{\alpha}\adj e^{-\hat{A}}\hat{a}_i e^{\hat{A}}|\pz}|^2\nonumber.
\end{align}
For computational feasibility, we take only the leading term of $e^{-\hat{A}}\hat{a}_i e^{\hat{A}} = \hat{a}_i + [\hat{a}_i,\hat{A}] + \ldots$, resulting in the working expression for the spectroscopic factor:
\begin{equation}
	S_{\alpha}\approx\sum_{i}^{\bvec{H}}|\braket{\pz|\hat{\mathcal{R}}_{\alpha}\adj\hat{a}_i|\pz}|^2.
\end{equation}

\subsection{Computational scaling}
The computational scaling of the IP-EOM-DSRG method is determined by the cost of the underlying MR-DSRG method and the cost of the IP-EOM-DSRG calculation itself.
In the common situation of $N_{\mathbf{V}}>N_{\mathbf{C}}\gg N_{\mathbf{A}}$, DSRG-MRPT2 scales as $\mathcal{O}(N_{\mathbf{C}}^2N_{\mathbf{V}}^2)$, and DSRG-MRPT3 and MR-LDSRG(2) scale as  $\mathcal{O}(N_{\mathbf{C}}^2N_{\mathbf{V}}^4)$, with the latter requiring an iterative procedure. 
The IP-EOM step scales as $\mathcal{O}(N_{\mathbf{C}}^3N_{\mathbf{V}}^2)$, arising from the cost of contracting the $\mathbf{CCVV}$ block of $\barh$ with the $\hat{a}^{\mathbf{V}}_{\mathbf{CC}}$ operators.
The IP-EOM-DSRG scaling with respect to the active space size is of $\mathcal{O}(N_{\mathbf{A}}^{10})$, arising from the contractions between the $\mathbf{AAAA}$ block of $\barh$ with the $\hat{a}^{\mathbf{A}}_{\mathbf{AA}}$ operators.

\section{Computational details}
\label{sec:comp}
The IP-EOM-DSRG method has been implemented in \textsc{NiuPy}, a library of multireference excited state methods written in Python.\cite{niupy}
The library uses a development branch of the \textsc{Wick\&d} library for the automatatic derivation of spin-integrated multireference many-body theories.\cite{evangelista.2022.10.1063/5.0097858}
All systems considered in this work are sufficiently small such that the full matrix $\bar{\bvec{H}}'$ can be held in memory and directly diagonalized.
Alternatively, the code has the ability to build the matrix-vector products $\bar{\bvec{H}}'\bvec{r}'_{\alpha}$ directly from the tensor elements of the (orthogonalized) effective Hamiltonian, instead of storing the full matrix $\bar{\bvec{H}}'$, which can be straightforwardly interfaced with most Hermitian iterative eigensolvers such as the Davidson--Liu algorithm.\cite{davidson.1975.10.1016/0021-99917590065-0,liu1978simultaneous}
All CASSCF and MR-DSRG calculations are performed with a development branch of the \textsc{Forte} quantum chemistry package.\cite{evangelista.2024.10.1063/5.0216512}
Unless otherwise stated, all electrons are correlated.
The choices of active space for all calculations, along with CASSCF reference energies, where applicable, can be found in Sec.~I of the Supporting Information.
All IP-EOM-DSRG calculations make use of the sequential orthogonalization scheme by K{\"o}hn and coworkers,\cite{kohn.2020.10.1080/00268976.2020.1743889} with thresholds of $\eta_1=10^{-5}$ and $\eta_2=10^{-10}$ unless otherwise stated.
All plots have been generated with the \textsc{matplotlib} and \textsc{seaborn} packages.\cite{hunter.2007.10.1109/MCSE.2007.55, waskom.2021.10.21105/joss.03021}
The conversion factors of $1$ Hartree $=27.211\ 386$ eV and $1$ Hartree $=2.194\ 746\times10^5$ cm$^{-1}$ are used throughout this work.\cite{tiesinga.2021.10.1063/5.0064853}

\section{Results and discussion}
\label{sec:results}
\subsection{Vertical ionization energies}
\label{sec:vert_ip}
The most common application of IP-EOM methods is predicting vertical ionization energies.
To test the accuracy of IP-EOM-DSRG, we have computed the lowest few vertical ionization energies of small molecules containing first-row elements (\ch{HF}, \ch{H2O}, \ch{CO}, \ch{N2}, \ch{F2}, \ch{CS}, \ch{C2H4}, and \ch{H2CO}) at their respective equilibrium geometries, comparing to extrapolated semistochastic heat-bath CI (SHCI)\cite{holmes.2017.10.1063/1.4998614}.
To enable a direct comparison, the molecular geometries, choice of basis, and the number of excited states computed for each system are identical to those of Chatterjee and Sokolov,\cite{chatterjee.2020.10.1021/acs.jctc.0c00778} which considered MR-ADC(2)-X and also includes data from SHCI, SR-ADC(2/3), and EOM-CCSD.
The equilibrium geometries are taken from Trofimov and Schirmer,\cite{trofimov.2005.10.1063/1.2047550} and the stretched geometries are obtained by doubling the bond lengths while keeping the bond angles fixed.
For \ch{C2H4} and \ch{H2CO}, this involves stretching the C--C and C--O bonds, respectively.
The aug-cc-pVDZ\cite{dunning.1989.10.1063/1.456153,kendall.1992.10.1063/1.462569,woon.1993.10.1063/1.464303} basis set is used for all atoms except for the hydrogen in \ch{C2H4} and \ch{H2CO}, where the cc-pVDZ basis set is used.\cite{dunning.1989.10.1063/1.456153}
EOM-CCSDT calculations are carried out in the \textsc{CCpy} package.\cite{ccpy}
The EOM-DSRG-PT2 calculations employ a flow parameter of $s=0.5$ $E_{\mathrm{h}}^{-2}$, and the EOM-DSRG-PT3 and EOM-DSRG(2) calculations employ a flow parameter of $s=1.0$ $E_{\mathrm{h}}^{-2}$.
The choices of flow parameters are in line with the recommendation from a previous benchmarking work on the optimal flow parameters for excited states in the MR-DSRG framework.\cite{wang.2023.10.1021/acs.jctc.2c00966}
We note here that the flow parameters chosen are not necessarily optimal for this particular test set, but are selected to reflect the typical values used in the literature.
In fact, there are choices of $s$ for PT2 ($\approx0.25$ $E_{\mathrm{h}}^{-2}$) and PT3 ($\approx0.5$ $E_{\mathrm{h}}^{-2}$) that will result in more compact error distributions for this test set. 
We will discuss the dependence of the vertical ionization energies on the flow parameter in more detail in \cref{sec:flow_param}.
The raw data used in this section is provided in Figs.~S1 and S2 of the Supporting Information.

\begin{figure*}[!htb]
	\centering
	\includegraphics[width=6.25in]{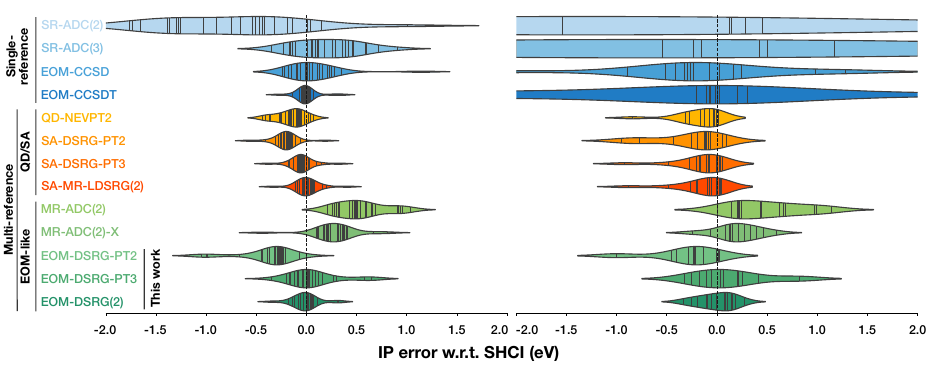}
	\caption{Left: violin plots (kernel density estimate of the error distribution, with data points represented as sticks within) of the errors of 25 vertical excitation energies at equilibrium geometries of \ch{HF}, \ch{H2O}, \ch{CO}, \ch{N2}, \ch{F2}, \ch{CS}, \ch{C2H4}, and \ch{H2CO}, computed with selected theoretical methods, compared to extrapolated SHCI. Right: violin plots of the errors of 14 vertical excitation energies for \ch{HF}, \ch{H2O}, \ch{N2}, \ch{F2}, \ch{C2H4}, and \ch{H2CO} at stretched geometries. A flow parameter of $s=1.0$ $E_{\mathrm{h}}^{-2}$ is used for the EOM-DSRG(2) and EOM-DSRG-PT3 calculations, and $s=0.5$ \sunit is used for EOM-DSRG-PT2 (see text for details).}
	\label{fig:ip-eqm-str}
\end{figure*}

\begin{figure}[!htb]
	\centering
	\includegraphics[width=3.125in]{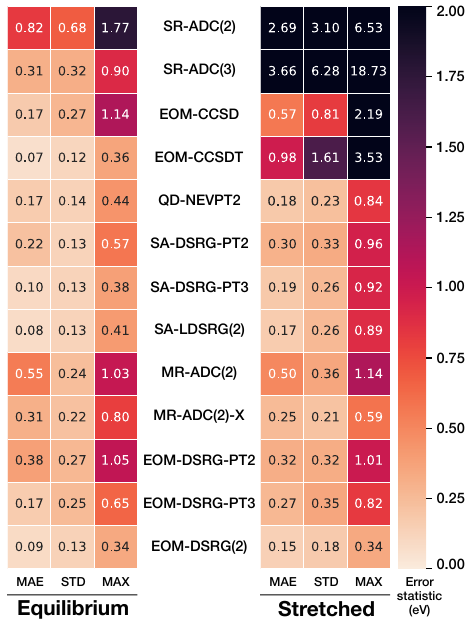}
	\caption{Summary statistics of the errors (w.r.t. SHCI) of the vertical ionization energies computed with various methods. The mean absolute error (MAE), standard deviation (STD), and maximum absolute error (MAX) are shown for each method.}
	\label{fig:error-heatmap}
\end{figure}

We report the error distribution for all methods considered in \cref{fig:ip-eqm-str} and summary error statistics in \cref{fig:error-heatmap}.
Considering \cref{fig:ip-eqm-str}, we can see that at equilibrium bond lengths, the most rigorous DSRG approach [IP-EOM-DSRG(2)] predicts accurate vertical ionization potentials all molecules, with an error profile very similar to that of IP-EOM-CCSDT---the former having a mean absolute error ($\mathrm{MAE}=N^{-1}\sum_i^N |\Delta\omega_i|$) of 0.09 eV, close to the 0.07 eV value achieved by the latter (see \cref{fig:error-heatmap}).
This is with the IP-EOM-DSRG formalism retaining up to 2h1p-type excitations, whereas IP-EOM-CCSDT includes 3h2p-type excitations at a significantly higher computational scaling of $\mathcal{O}(N_{\mathbf{C}}^3N_{\mathbf{V}}^4)$.
The IP-EOM-DSRG-PT2 method underestimates the excitation energies, with a larger MAE of 0.38 eV, whereas the IP-EOM-DSRG-PT3 method has a smaller systematic error, reflected in an MAE of 0.17 eV, only slightly higher than that of IP-EOM-DSRG(2).
However, both IP-EOM-DSRG-PT2 and PT3 have a higher spread in the error distribution, reflected in the larger standard deviation (STD) of the errors of 0.27 eV and 0.25 eV, compared to 0.13 eV for IP-EOM-DSRG(2).
From these observations already, we can see the systematic improvement brought about by improving the underlying description of electron correlation, roughly following the trend of IP-EOM-DSRG(2) $\geq$ IP-EOM-DSRG-PT3 $\gg$ IP-EOM-DSRG-PT2.

At stretched geometries, the same trends persist, with IP-EOM-DSRG(2) achieving a similarly compact and centered error profile with respect to SHCI, albeit at a slightly larger mean absolute error.
The IP-EOM-DSRG(2) method outperforms all other methods in MAE, STD, and maximum errors (MAX), whereas IP-EOM-DSRG-PT2/3 are comparable to the MR-ADC(2) method,\cite{chatterjee.2019.10.1021/acs.jctc.9b00528} while being significantly better than all other single-reference methods considered, most of which have trouble with even qualitatively correct descriptions of the target states.
Notably, the IP-EOM-CCSDT results become significantly worse at stretched geometries, even worse than the IP-EOM-CCSD.

The poor performance of IP-EOM-CCSDT at stretched geometries deserves some attention, as it illustrates a more subtle point when comparing the accuracies of excitation energies.
The large error spread in this case is primarily driven by two IPs of the \ch{N2} molecule at $2.196$ \AA\ bond length.
Although neither CCSD nor CCSDT can accurately describe the triple bond-breaking process in \ch{N2}, the CCSDT ground state potential energy curve (PEC) falls below the variational limit more severely than the CCSD PEC. 
As a consequence, the ionization potential is overestimated in CCSDT.
Multireference methods, such as EOM-DSRG and MR-ADC, do not suffer from this unbalanced description of reference and target states, as their respective reference states can qualitatively describe processes involving multiple bonds breaking, thereby avoiding overestimating IPs.

\begin{figure}[!htb]
	\centering
	\includegraphics[width=3.125in]{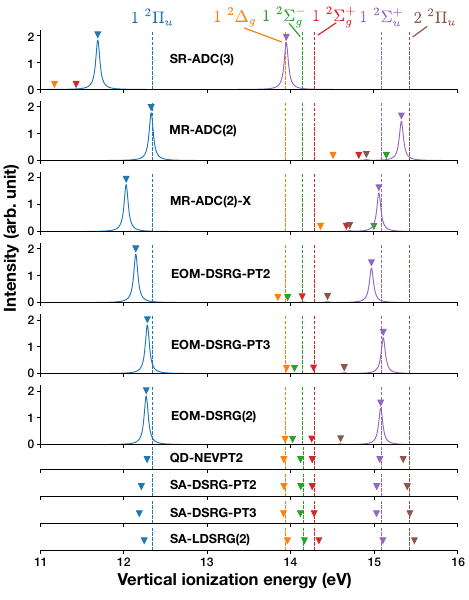}
	\caption{Simulated photoelectron spectrum of \ch{C2} at its equilibrium geometry computed with various theoretical methods. The reference SHCI results are shown as dotted vertical lines. The vertical ionization energies from other theoretical methods are shown as peaks broadened by a Lorentzian function with a half-width at half-maximum of 0.03 eV. The intensities are given directly by the spectroscopic factors. Triangles indicate locations of transitions, even if their intensity is too low to be visible. For reference, results from select QD/SA methods are provided at the bottom of the figure without intensity information.} 
	\label{fig:c2}
\end{figure}

Next, we consider the carbon dimer (not contained in the previous dataset), an archetypal system with strong multireference character, even at its equilibrium geometry, which can compromise the computation of ionized states due to significant differential correlation effects from the ground state.
In \cref{fig:c2} we report the vertical ionization energies and spectroscopic factors of the low-lying electronic states of \ch{C2} at its equilibrium geometry (1.2425\ \AA), computed with EOM-DSRG-PT2/3 and EOM-DSRG(2) methods, and compared to results from SR-ADC(3), MR-ADC(2) and MR-ADC(2)-X methods,\cite{chatterjee.2020.10.1021/acs.jctc.0c00778} along with quasi-degenerate (QD) NEVPT2\cite{angeli.2004.10.1063/1.1778711} (reproduced from Ref.~\citenum{chatterjee.2019.10.1021/acs.jctc.9b00528}) and state-averaged (SA) DSRG-MRPT2/3 and MR-LDSRG(2) results.
The differential correlation effects can be most clearly seen in the $2\ ^2\Pi\ung$ state (brown in \cref{fig:c2}), whose dominant configuration can be obtained from the \ch{C2} ground state by an ionization followed by a double excitation ($(2\sigma\ung)^2(1\pi\ung)^4(3\sigma\ger)^0\rightarrow(2\sigma\ung)^2(1\pi\ung)^1(3\sigma\ger)^2$), \textit{i.e.}, it has significant 3h2p-character. 
None of the EOM-like methods are able to even qualitatively describe the state, as they all make use of up to 2h1p operators, whereas all QD/SA methods are able to describe the state accurately, due to the explicit consideration of the target state in the calculations.
This may be a case where the state-transfer internal operators in MR-ADC(2)(-X)\cite{chatterjee.2019.10.1021/acs.jctc.9b00528} could be used to improve the description of the state, since all of the involved orbitals are in the CAS.
However, the MR-ADC(2)(-X) methods exhibit a similar failure to describe this state. 
This might point to the need for external 3h2p operators to capture orbital relaxation and correlation effects in this case, or it could be that the relevant state-transfer operators were not captured in the MR-ADC(2)(-X) calculations.

\subsection{Dependence on the flow parameter}
\label{sec:flow_param}

\begin{figure}[!htb]
	\centering
	\includegraphics[width=3.125in]{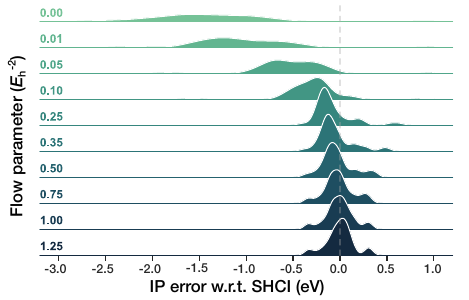}
	\caption{The dependence of the vertical ionization energies computed with IP-EOM-DSRG(2) on the flow parameter $s$.}
	\label{fig:flowparam}
\end{figure}

In this section, we examine how the flow parameter value affects the error statistics of the IP-EOM-DSRG methods.
In \cref{fig:flowparam}, we show the dependence of all IP-EOM-DSRG(2) vertical ionization energies on the flow parameter $s$ for the small molecules considered discussed in \cref{sec:vert_ip}, with the exception of \ch{C2}.
In the $s=0$ \sunit case, the bare CASSCF Hamiltonian is used, and the IP-EOM-DSRG(2) method reduces to an EOM method based on a fully internally contracted MRCI method (fic-MRCI).
As can be seen from the figure, the performance of the bare Hamiltonian is very poor, with large systematic errors ($\mathrm{MAE} = 1.43$ eV) and a large spread ($\mathrm{STD} = 0.69$ eV) in the error distribution.
As $s$ is increased, the errors quickly become smaller and the distribution more compact, achieving good accuracy in a large range of $s$ values (0.25--1.25 \sunit).
This can be understood as the result of the DSRG transformation folding in dynamic correlation effects into the effective Hamiltonian, which makes the compact excitation operator manifold more effective as the renormalization of the Hamiltonian proceeds.
However, MR-LDSRG(2) starts to encounter convergence issues at $s>1.25$ \sunit for some systems, which could mean the MR-LDSRG(2) operator equation [\cref{eq:manybody-cond}] has no solution at those values of $s$.
Analogous figures for EOM-DSRG-PT2/3 are provided in Figs.~S3 and S4 in the Supporting Information.

\begin{figure}[!htb]
	\centering
	\includegraphics[width=3.125in]{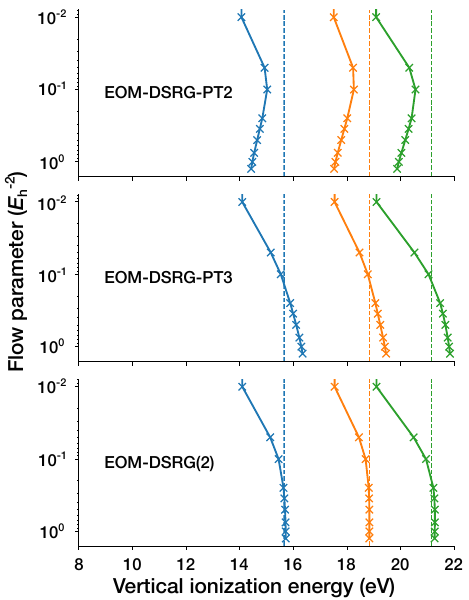}
	\caption{The dependence of the vertical ionization energies of \ch{F2} at its equilibrium geometry on the flow parameter $s$.}
	\label{fig:root_f2}
\end{figure}

The advantage of the iterative MR-LDSRG(2) method over the perturbative methods is highlighted in \cref{fig:root_f2}, where we show the dependence of the vertical ionization energies of \ch{F2} at its equilibrium geometry on the flow parameter $s$.
Here, we can see that the EOM-DSRG(2) IPs stably converge to the SHCI values as $s$ is increased, whereas the EOM-DSRG-PT2 IPs come close but then diverge, and the EOM-DSRG-PT3 IPs cross the SHCI values but then overshoot.
Plots of the $s$-dependence for every system considered in the test set are provided in Figs.~S5 and S6 in the Supporting Information.

\begin{figure}[!htb]
	\centering
	\includegraphics[width=3.125in]{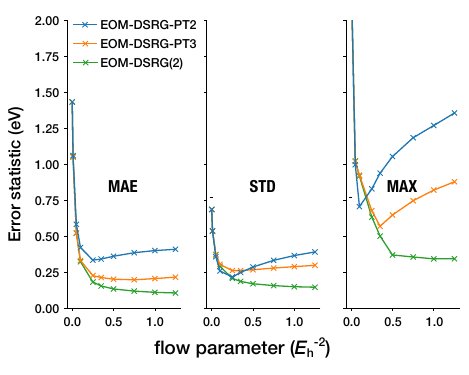}
	\caption{Summary statistics of the errors (w.r.t. SHCI) of the vertical ionization energies as a function of the flow parameter $s$ for EOM-DSRG methods. The mean absolute error (MAE), standard deviation (STD), and maximum absolute error (MAX) are shown for each method.}
	\label{fig:flowparam-error}
\end{figure}

The observed flow parameter trends for the three truncated schemes is further borne out in \cref{fig:flowparam-error}, where we show the mean absolute error (MAE), standard deviation (STD), and maximum absolute error (MAX) of the vertical ionization energies computed with EOM-DSRG-PT2/3 and EOM-DSRG(2) as a function of the flow parameter $s$.
Consistent with the findings of Ref.~\citenum{wang.2023.10.1021/acs.jctc.2c00966}, the error profiles of PT2 as a function of $s$ exhibit deep and unstable minima at small $s$ near $0.5$ \sunit, whereas the error profiles of PT3 and MR-LDSRG(2) has shallower minima or even flat profiles.
The larger computational prefactor and convergence difficulties at large $s$ for the MR-LDSRG(2), together with the results in this section, lead us to tentatively recommend the IP-EOM-DSRG-PT3 method with $s\in[0.5,1.0]$ \sunit for the best balance between accuracy and computational cost.

\subsection{Spectroscopic constants}
\label{sec:spectroscopic_constants}
\begin{figure*}[!htb]
	\centering
	\includegraphics[width=6.25in]{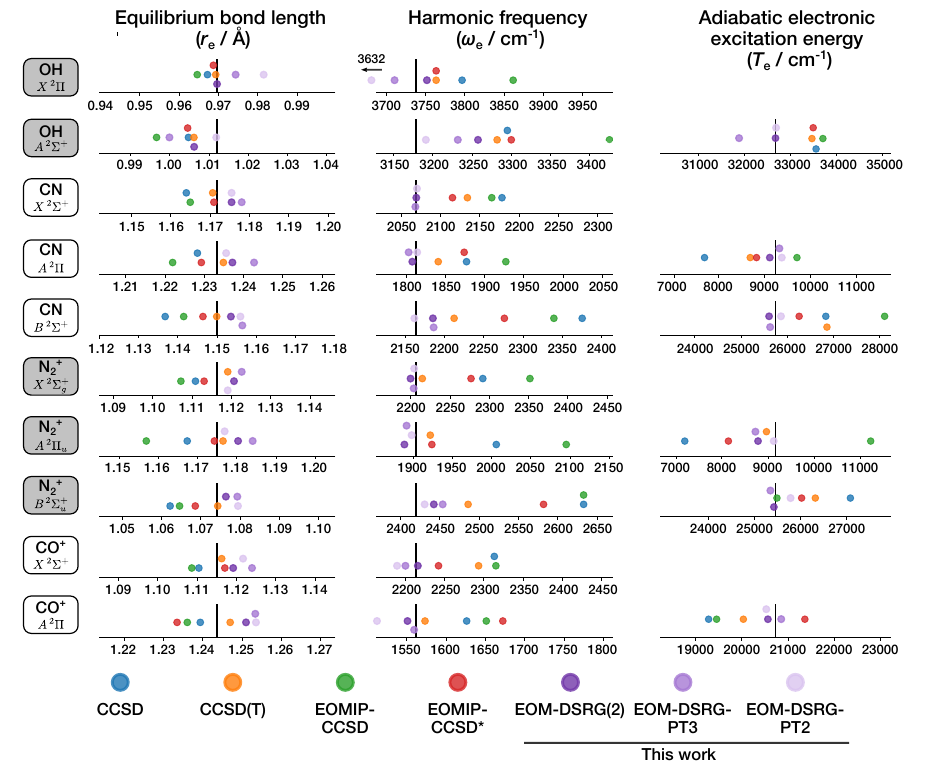}
	\caption{Spectroscopic constants for select electronic states of \ch{OH}, \ch{CN}, \ch{N2+}, and \ch{CO+} radicals, compared to the experimental values (black vertical lines) from Huber and Herzberg.\cite{huber.1979.10.1007/978-1-4757-0961-2} The adiabatic transition energies are from the electronic ground states of the respective radicals.}
	\label{fig:spectroscopic_constants}
\end{figure*}

\begin{figure}[!htb]
	\centering
	\includegraphics[width=3.125in]{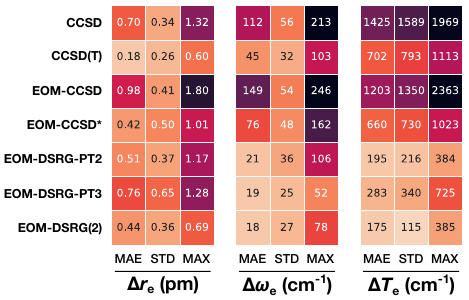}
	\caption{Summary error statistics for select electronic states of \ch{OH}, \ch{CN}, \ch{N2+}, and \ch{CO+} radicals relative to experimental values. The mean absolute error (MAE), standard deviation (STD), and maximum absolute error (MAX) are shown for each method.}
	\label{fig:spec_error}
\end{figure}

To assess the ability of the EOM-DSRG methodology to accurately determine the spectroscopic properties for multiple target states simultaneously, we compute the spectroscopic constants of a sample of radical species, starting from their respective closed-shell electron-attached counterparts.
In \cref{fig:spectroscopic_constants}, we compare IP-EOM-DSRG spectroscopic constants to those from IP-EOM-CCSD and IP-EOM-CCSD* (with the latter including 3h2p determinants approximately)\cite{saeh.1999.10.1063/1.480171} and CCSD and CCSD(T) based on UHF or quasi-restricted HF references.\cite{rittby.1987.10.1021/j100322a004}
The MAE, STD, and MAX of the spectroscopic constants are shown in \cref{fig:spec_error}.
We have used the local interpolating moving least squares method by Bender and coworkers\cite{bender.2014.10.1063/1.4862157} as implemented in the \textsc{Psi4} package\cite{smith.2020.10.1063/5.0006002} to obtain the spectroscopic constants.
Grids of 31 points spaced at $0.001$ \AA\ around the equilibrium bond lengths are used for the fitting.
The cc-pVTZ basis set\cite{dunning.1989.10.1063/1.456153} is used for all atoms in these calculations for comparison with previous theoretical values computed by Saeh and Stanton.\cite{saeh.1999.10.1063/1.480171}
The raw data for this section is provided in Tables~S1-S4 in the Supporting Information.

We note that state-specific approaches like CCSD and CCSD(T) in \cref{fig:spectroscopic_constants} require a separate calculation for each state.
State-specific approaches forgo orthogonality among the states, and in return usually give more accurate properties for the target states, as they are individually optimized.
If the target states are not the ground states of their respective symmetry sectors, methods such as $\Delta$-CC\cite{lee.2019.10.1063/1.5128795,damour.2024.10.1021/acs.jctc.4c00034} or excited-state-specific CC\cite{tuckman.2023.10.1021/acs.jctc.3c00194} may be required.

We can observe that all three IP-EOM-DSRG methods perform well for the states considered, and generally produces spectroscopic constants that are in good agreement with the experimental values with a moderately large basis set.
All methods achieve similar performance for the equilibrium bond lengths ($r_{\mathrm{e}}$).
This is expected as all methods considered here are able to capture weak correlation effects typically present around equilibrium geometries.
This property can also be thought of as depending on the first derivative of the energy, and is therefore less sensitive to the choice of method.
However, already for the harmonic frequencies, we can see, in the middle panel of \cref{fig:spec_error}, a marked reduction in the MAEs from the order of 100 cm$^{-1}$ for single-reference methods to the order of 10 cm$^{-1}$ for the IP-EOM-DSRG methods, with single-reference methods overestimating the harmonic frequencies (\textit{i.e.}, predicting stiffer bonds), as they systematically undercorrelate the states away from their equilibrium geometries.
This property depends on the second derivative of the energy, and therefore demands a more accurate description of the PES. 
The utility of a multi-reference treatment is most clearly reflected in the adiabatic electronic excitation energies (right panels of \cref{fig:spectroscopic_constants,fig:spec_error}), where the IP-EOM-DSRG methods predict the much more accurate gaps between the ionized states than the single-reference methods, following the trend observed in \cref{sec:vert_ip}.
IP-EOM-DSRG-PT2/3 results generally compare well with those from IP-EOM-DSRG(2), but can become less reliable for certain states. See for example the harmonic frequencies for the $\statex\ ^2\Pi$ state of \ch{OH^.}. The perturbative schemes sometimes appear to perform better than IP-EOM-DSRG(2), but this is likely due to error cancellation as seen, for example, in the harmonic frequencies of $\statea\ ^2\Sigma^+$ state of \ch{OH^.}, where increasing the level of theory causes a rightward shift that corresponds to larger errors.

\subsection{Potential energy surfaces}
\begin{figure}[!htb]
	\centering
	\includegraphics[width=3.125in]{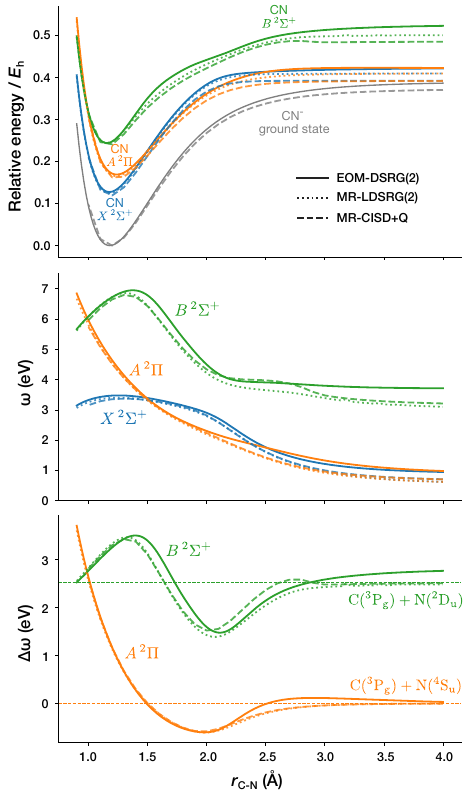}
	\caption{Potential energy surfaces of the $\statex\ ^2\Sigma^+$, $\statea\ ^2\Pi$ and $\stateb\ ^2\Sigma^+$ states of the cyanide radical. In the top figure, ``Relative energy'' denotes the energies of the radical states relative to the minimum energy of the \ch{CN-} ground state computed by state-specific MR-LDSRG(2) for EOM-DSRG and SA-DSRG, and state-specific fic-MRCISD+Q for fic-MRCISD+Q. In the middle figure, $\omega$ denotes the vertical ionization energies of the radical states from the \ch{CN-} ground state. In the bottom figure, $\Delta \omega$ denotes the spacings between the $\statea\ ^2\Pi$ and $\stateb\ ^2\Sigma^+$ states to the $\statex\ ^2\Sigma^+$ states. The dotted lines indicate the FCI dissociation limits of the corresponding states. The position of the green dotted line is computed as the FCI/cc-pVTZ energy difference between the $^2\mathrm{D}\ung$ and $^4\mathrm{S}\ung$ states of the nitrogen atom.} 
	\label{fig:cn_curve}
\end{figure}

A key advantage of the MR-DSRG formalism is its ability to generate smooth and qualitatively correct potential energy surfaces due to its intruder-free property and its use of MCSCF reference wavefunctions.
The EOM-DSRG formalism is then expected to also be intruder-free, and consequently avoid discontinuities due to the presence of intruders.
However, the use of truncation thresholds in the orthogonalization procedure may introduce small discontinuities in the potential energy surfaces when there are large changes in the number of orthogonalized operators.\cite{li.2025.10.1063/5.0261000}.
The thresholds we use throughout this work are small enough that all PESes are practically continuous.
As an example, we consider the potential energy curves of the $\statex\ ^2\Sigma^+$, $\statea\ ^2\Pi$ and $\stateb\ ^2\Sigma^+$ states of the cyanide radical.
The corresponding curves are shown in \cref{fig:cn_curve}, where we report IP-EOM-DSRG results, and compare them to those obtained via the state-averaged MR-LDSRG(2)\cite{li.2018.10.1063/1.5019793} and fully internally contracted MRCISD with Davidson correction (fic-MRCISD+Q)\cite{sivalingam.2016.10.1063/1.4959029} potential energy surfaces, computed with the \textsc{Forte}\cite{evangelista.2024.10.1063/5.0216512} and \textsc{ORCA}\cite{neese.2020.10.1063/5.0004608} packages respectively.

Looking at the top panel of \cref{fig:cn_curve} (energies relative to the minimum on the \ch{CN-} curve), we can see that the IP-EOM-DSRG(2) method produces smooth potential energy surfaces for all three states, and they are qualitatively correct compared to the fic-MRCISD+Q potential energy surfaces, which we consider to be the most accurate reference in this case.
We also note that the correct asymptotic degeneracies are obtained with the EOM-DSRG methods.
Focusing in the middle figure of \cref{fig:cn_curve} (vertical ionization energies), we can see that the IP-EOM-DSRG(2) method produces curves that are largely parallel to the fic-MRCISD+Q, except for a marked deviation in the $\stateb\ ^2\Sigma^+$ state at around $2.8$ \AA, where an avoided crossing occurs.\cite{terashkevich.2021.10.1016/j.jqsrt.2021.107916}
An avoided crossing is present in the IP-EOM-DSRG(2) PES, but occurs at around $2.2$ \AA.
This could be due to IP-EOM-DSRG(2) not having enough flexibility to capture the full differential correlation effects present in the $\stateb\ ^2\Sigma^+$ statedue to significant recoupling of the electrons in this region of the curve.
The bottom figure of \cref{fig:cn_curve} (excitation energies of the ionized state) reveals another limitation of the IP-EOM-DSRG(2) method compared to state-averaged methods like MR-LDSRG(2) and fic-MRCISD+Q: the $\stateb\ ^2\Sigma^+$ state at dissociation is not well described by IP-EOM-DSRG(2), although its zeroth-order reference state is qualitatively correct.
We would expect IP-EOM-DSRG(2) to more accurately describe the dissociation limit, as the $\mathrm{N}(^2\mathrm{D}\ung)$ atomic state is dominated by 1h excitations from the \ch{CN-} ground state, and all nearby $\Sigma^+$ states are dominated by at most 2h1p excitations.
The comparatively better description of the potential energy curves of \ch{CN^.} afforded by the MR-LDSRG(2) and fic-MRCISD+Q methods is then attributed to the double excitations in the $(N-1)$ electron Hilbert space that these methods introduce.
This result suggests that a route to improve the accuracy of the \ch{CN^.} results is by inclusion of the 3h2p excitations in the IP-EOM-DSRG, which are effectively treated in the MR-LDSRG(2) and fic-MRCISD+Q methods.

We show the same potential energy surfaces for IP-EOM-DSRG-PT2/3, compared to IP-EOM-DSRG(2) in Fig.~S7 in the Supporting Information.
The IP-EOM-DSRG-PT3 curves almost exactly overlap with the IP-EOM-DSRG(2) curves, while IP-EOM-DSRG-PT2 slightly over-correlates, \textit{i.e.}, underestimates ionization energies for all states.
In increasing the level of theory from IP-EOM-DSRG-PT2, through IP-EOM-DSRG-PT3, to IP-EOM-DSRG(2), the FCI dissociation limit of the $\stateb\ ^2\Sigma^+$ state is improved.

\subsection{Size intensitivity}
\label{sec:size_intensivity}
As mentioned in \cref{sec:theory}, the IP-EOM-DSRG formalism is not rigorously size intensive since the resulting $\bar{H}$ does not satisfy projective conditions.
In most applications, it is paramount that deviations from core intensivity are significantly smaller than the target excitation energies.
Core intensivity describes the invariance of the ionization potential with respect to the addition of non-interacting core and virtual orbitals, while the active orbitals remain unchanged.
This quantity reflects the most common situation in which the system of interest is studied in a variety of larger, weakly interacting environments.
In \cref{tab:core_intensivity} we quantify the core-intensivity errors of the IP-EOM-DSRG formalism by computing the errors in the first vertical ionization energies of \ch{HF} at the equilibrium and stretched geometries, in the presence of a variable number of non-interacting helium atoms (spaced $10000$ \AA\ apart), compared to an isolated \ch{HF} molecule.
We use the aug-cc-pVDZ basis and obtain $\barh$ from the DSRG-MRPT3 with flow parameters of $0.5$, $1$, and $10$ \sunit.
A consistent active space of 6 electrons in 5 orbitals (H 1s, F 2s/2p orbitals) is used for all calculations.
\begin{table}[!htb]
\footnotesize
\renewcommand*{\arraystretch}{1.5}
\begin{tabular}{ccccccc}
	\hhline{=======}
	\multicolumn{7}{c}{Core-intensivity errors (meV) = IP(HF$\cdots$ He$_n$) - IP(HF)} \\ \hline
\multirow{2}{*}{\ch{HF} + $N$ \ch{He}} & \multicolumn{2}{c}{$s=0.5$ \sunit} & \multicolumn{2}{c}{$s=1$ \sunit} & \multicolumn{2}{c}{$s=10$ \sunit}\\
& Equilibrium & Stretched & Equilibrium & Stretched & Equilibrium & Stretched \\ \hline
1 & $-0.147$  & $-0.077$ & $-0.058$ & $-0.031$ & $0.000$ & $0.000$ \\
2 & $-0.294$  & $-0.153$ & $-0.115$ & $-0.061$ & $0.000$ & $0.000$ \\
3 & $-0.440$  & $-0.230$ & $-0.173$ & $-0.092$ & $0.000$ & $0.000$ \\
4 & $-0.587$  & $-0.307$ & $-0.231$ & $-0.122$ & $0.000$ & $0.000$ \\
5 & $-0.734$  & $-0.383$ & $-0.289$ & $-0.153$ & $0.000$ & $0.000$ \\ \hline
0 & $16.183$  & $12.924$ & $16.348$ & $12.928$ & $16.743$&$12.972$ \\
\hhline{=======}
\end{tabular}
\caption{Core-intensivity errors (in meV) of the IP-EOM-DSRG-PT3 for \ch{HF} at equilibrium and stretched geometries in the presence of $N$ non-interacting helium atoms. The equilibrium geometry is at $0.917$ \AA\ and the stretched geometry is at $1.834$ \AA. The vertical ionization energies (in eV) without any helium atoms are shown in the last row.}
\label{tab:core_intensivity}
\end{table}
For the smaller $s$ values we can see that the core-intensivity errors are typically minuscule (in the order of 0.1 meV for $s$ = 0.5 \sunit) compared to the systematic errors of the IP-EOM-DSRG formalism (in the order of 100 meV), and are largely independent of the bond length, meaning that spectral gaps are even less affected by the core-intensivity errors.
Going from $s$ = 0.5 to 1.0 \sunit, the error due to an additional helium atom goes from about 0.150 to 0.060 meV at the equilibrium geometry, and from about 0.080 meV to 0.030 meV at the stretched geometry, improving the core-extensivity error significantly.
When $s$ is increased to 10.0 \sunit, the core-intensivity errors essentially vanish (all values are below $3\times 10^{-7}$ eV).
This can be understood as follows: the addition of non-interacting helium atoms only contributes more core-virtual operators, and the corresponding projective conditions are satisfied in the limit of $s\rightarrow\infty$;\cite{li.2019.10.1146/annurev-physchem-042018-052416} hence, the IP-EOM-DSRG formalism is expected to be rigorously core-intensive in this limit.

We also investigate the full intensivity of the IP-EOM-DSRG formalism by computing the IP of two non-interacting \ch{HF} molecules at its equilibrium and stretched geometries, with the same basis set and active space as above.
This tests the invariance of the IP with respect to the addition of non-interacting orbitals in all three orbital partitions (core, active, and virtual).
Here, the corresponding projective conditions cannot be satisfied even in the limit of $s\rightarrow\infty$.\cite{li.2019.10.1146/annurev-physchem-042018-052416}
For a flow parameter of $s=0.5$ \sunit, the IP-EOM-DSRG-PT3 method gives a full intensivity error of $34.9$ meV at the equilibrium geometry, and $16.2$ meV at the stretched geometry, which is significantly larger than the core-intensivity errors, but still around an order of magnitude smaller than the systematic errors of the method.

\section{Conclusions}
\label{sec:conclusions}
In this work, we presented the first formulation of a multireference equation-of-motion method based on the DSRG formalism for computing ionization potentials.
Three methods are derived from this formalism, namely IP-EOM-DSRG(2), based on the iterative MR-LDSRG(2) scheme, and IP-EOM-DSRG-PT2/3, based on the perturbative DSRG-MRPT2/3 approaches.
These methods were benchmarked on near-FCI quality vertical ionization energies for a set of small molecules at both their equilibrium and stretched geometries; the spectroscopic constants of low-lying electronic states of several radical species; and the potential energy surfaces of the $\statex\ ^2\Sigma^+$, $\statea\ ^2\Pi$ and $\stateb\ ^2\Sigma^+$ states of the cyanide radical.
All three methods were found to produce accurate spectroscopic quantities, such as ionization potentials and various spectroscopic constants, even in systems with strongly correlated ground states.
IP-EOM-DSRG(2) outperforms other single- and multi-reference EOM-like methods, and is even competitive with some state-averaged/quasi-degenerate methods that work directly in the $(N-1)$ electron Hilbert space.
IP-EOM-DSRG-PT3 was found to be almost as accurate as IP-EOM-DSRG(2) while being more computationally efficient, as it does not require the ground state amplitudes to be iteratively optimized.
It can also avoid the convergence issues that can sometimes occur in MR-LDSRG(2) calculations.
Both IP-EOM-DSRG(2) and IP-EOM-DSRG-PT3 yield accurate vertical ionization energies over a large range of flow parameters. The IP-EOM-DSRG-PT2 was instead found to be more sensitive to this choice and produced less accurate vertical ionization energies even at the optimal flow parameter.
The IP-EOM-DSRG methods were also found to produce smooth and qualitatively correct potential energy surfaces.
Issues surrounding missing orbital relaxation and correlation effects in IP-EOM-DSRG were discussed and attributed to missing 3h2p excitations in the present formulation.

Overall, the results presented in this work are encouraging and demonstrate the potential of the EOM-DSRG formalism and the utility of the IP-EOM-DSRG methods for accurately describing ionization potentials and other spectroscopic properties in molecular systems with strongly correlated ground states.
Work is already underway to extend the EOM-DSRG formalism for describing other spectroscopic processes such as core-ionization and electron excitation.
Extensions that account for spin-orbit coupling effects in the presence of strong correlation using the recently developed relativistic MR-DSRG formalism\cite{zhao.2024.10.1021/acs.jpclett.4c01372} will also be explored.
Future work should also focus on efficient, spin-adapted implementations of the present formalism, as well as the exploration of techniques to bypass the 4-body cumulant.
At a more fundamental level, one challenge to overcome is the formulation of a rigorously size-intensive IP-EOM-DSRG approach by enforcing the proper projective constraints.
For this, we intend to explore alternative formulations of the MR-DSRG theory that partially satisfies the projective condition, in the same vein as the pIC-MRCC approach.\cite{datta.2011.10.1063/1.3592494}

\section*{Author Information}
Zijun Zhao (ORCID: 0000-0003-0399-2968)\\
Shuhang Li (ORCID: 0000-0002-1488-9897)\\
Francesco A. Evangelista (ORCID: 0000-0002-7917-6652)

\section*{Supporting Information}
See the Supporting Information for additional details:
1) The choice of active spaces and reference energies for the CASSCF calculations, where applicable.
2) Tables of all vertical ionization energies computed with IP-EOM-DSRG(2), IP-EOM-DSRG-PT2, and IP-EOM-DSRG-PT3, compared with other methods.
3) Overall flow parameter dependence of the vertical ionization energies computed with IP-EOM-DSRG-PT2, and IP-EOM-DSRG-PT3.
4) Flow parameter dependence analysis for each of the ionized states considered in \cref{sec:vert_ip}.
5) Tables of spectroscopic constants discussed in \cref{sec:spectroscopic_constants}.
6) A zip file containing comma-separated value (CSV) files of the raw data for all figures in the main text and Supporting Information.

\section*{Acknowledgements}
This research was primarily supported by the U.S. Department of Energy under Award DE-SC0024532.
S.L. was supported to develop the \textsc{NiuPy} package by the National Science Foundation and the Molecular Sciences Software Institute under Grant No. CHE-2136142.
The authors thank Dr. Karthik Gururangan for help with the \textsc{CCpy} package, along with helpful discussions and comments on the manuscript.

\section*{Data availability}
The data that support the findings of this study are available within the article, its supporting information, and from the corresponding author upon reasonable request.


\begin{mcitethebibliography}{135}
\bibitem[Not()]{Note-1}
This is true assuming no truncation of the cluster operator. In
  single-reference EOM-CC, this can still hold true for truncated cluster and
  EOM operators, since determinants of different excitation ranks are
  orthogonal, which means for EOM-CCSD,
  $\bra{\Phi_0}\hat{\rho}_p\adj\hat{\rho}_q\hat{Q}_{\mathrm{TQ\dots}}=0$, where
  $\hat{Q}_{\mathrm{TQ\dots}}$ is the projector onto triple and higher
  excitations. In the MR case, decoupling of the model space and its orthogonal
  complement occurs if the lower excitations have been projected out of the
  higher excitations, which, in practice, is not done for excitation ranks
  higher than the truncation rank.\relax
\mciteBstWouldAddEndPunctfalse
\mciteSetBstMidEndSepPunct{\mcitedefaultmidpunct}
{}{\mcitedefaultseppunct}\relax
\EndOfBibitem
\end{mcitethebibliography}
\providecommand{\latin}[1]{#1}
\makeatletter
\providecommand{\doi}
  {\begingroup\let\do\@makeother\dospecials
  \catcode`\{=1 \catcode`\}=2 \doi@aux}
\providecommand{\doi@aux}[1]{\endgroup\texttt{#1}}
\makeatother
\providecommand*\mcitethebibliography{\thebibliography}
\csname @ifundefined\endcsname{endmcitethebibliography}
  {\let\endmcitethebibliography\endthebibliography}{}



\section*{Supplementary material}

**Supporting information: Multireference equation-of-motion driven similarity renormalization group: theoretical foundations and applications to ionized states**

Zijun Zhao,[1, a)] Shuhang Li,[1, b)] and Francesco A. Evangelista[1, c)]

*Department of Chemistry and Cherry Emerson Center for Scientific Computation, Emory University, Atlanta, Georgia 30322, USA*

[a)] Electronic mail: zijun.zhao@emory.edu
[b)] Electronic mail: shuhang.li@emory.edu
[c)] Electronic mail: francesco.evangelista@emory.edu

# I. CHOICE OF ACTIVE SPACE AND REFERENCE ENERGIES

Here we document the active space choices for the molecules studied in Section IV.A. The CASSCF reference energies are also reported. The active space is specified by the number of core and active orbitals per irrep, ordered in the 'Cotton ordering', i.e., [$a_g$, $b_{1g}$, $b_{2g}$, $b_{3g}$, $a_u$, $b_{1u}$, $b_{2u}$, $b_{3u}$] for the $D_{2h}$ point group, and [$a_1$, $a_2$, $b_1$, $b_2$] for the $C_{2v}$ point group. The atomic valence active space (AVAS) method[1] is used to select the initial active space orbitals as needed, and the AVAS subspace is specified for those cases.

| Molecule | Core | Active | AVAS | $E_{\text{CAS}}$ / $E_{\text{h}}$ |
|---|---|---|---|---|
| $C_2$ | [1,0,0,0,0,1,0,0] | [2,0,2,2,0,2,2,2] | | -75.652069959610 |
| $C_2H_4$ (eqm) | [1,0,0,0,0,1,0,0] | [3,0,1,2,0,3,2,1] | ["C(2s)","C(2p)","H(1s)"] | -78.994626688870 |
| $C_2H_4$ (str) | | | | -77.945136269290 |
| CO | [2,0,0,0] | [4,0,2,2] | | -112.885952774734 |
| CS | [4,0,1,1] | [4,0,2,2] | | -435.438613682326 |
| $F_2$ (eqm) | [1,0,0,0,0,1,0,0] | [2,0,1,1,0,2,1,1] | | -198.777678643099 |
| $F_2$ (str) | | | | -198.753977458044 |
| $H_2CO$ (eqm) | [2,0,0,0] | [5,0,2,3] | ["C(2s)", "C(2p)", "O(2s)", "O(2p)", "H(1s)"] | -114.014228155162 |
| $H_2CO$ (str) | | | | -113.767278864429 |
| $H_2O$ (eqm) | [1,0,0,0] | [4,0,2,3] | | -76.176803247525 |
| $H_2O$ (str) | | | | -75.884613320071 |
| HF (eqm) | [1,0,0,0] | [5,0,2,2] | | -100.173027089660 |
| HF (str) | | | | -100.006723519719 |
| $N_2$ (eqm) | [1,0,0,0,0,1,0,0] | [2,0,1,1,0,2,1,1] | | -109.108676505452 |
| $N_2$ (str) | | | | -108.785572038754 |

For the spectroscopic constants of radicals, the active space is specified as follows:

- $CN^-$: core: [2,0,0,0], active: [4,0,2,2], AVAS subspace: ["C(2s)", "C(2p)", "N(2s)", "N(2p)"]

- CO: same as above

- $N_2$: same as above

2

- OH$^-$: core: [1,0,0,0], active: [3,0,2,2]

For the CN binding curve, the same active space as CN$^-$ above is used.

All state-averaged calculations are performed using the same active space as above, using an ensemble of only the targeted states, with equal weights. Additional states of zero weight may be added to remedy the occasional root-flipping problem.

## II. VERTICAL IONIZATION ENERGIES OF MOLECULES AT EQUILIBRIUM GEOMETRIES

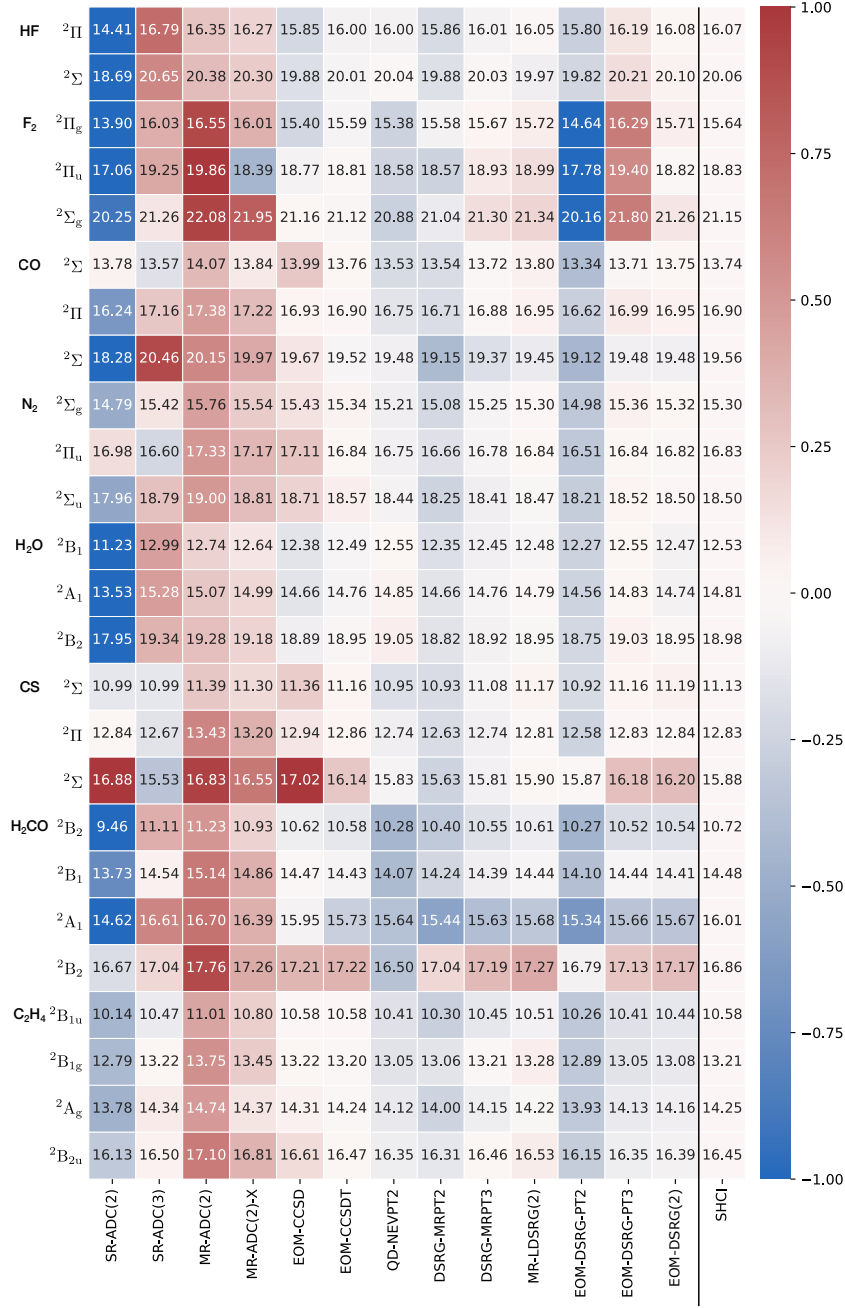

FIG. S1. Vertical ionization energies (IPs) of molecules at equilibrium geometries. The colors indicate the signed error (eV) of the IPs with respect to the SHCI values.

4

## III. VERTICAL IONIZATION ENERGIES OF MOLECULES AT STRETCHED GEOMETRIES

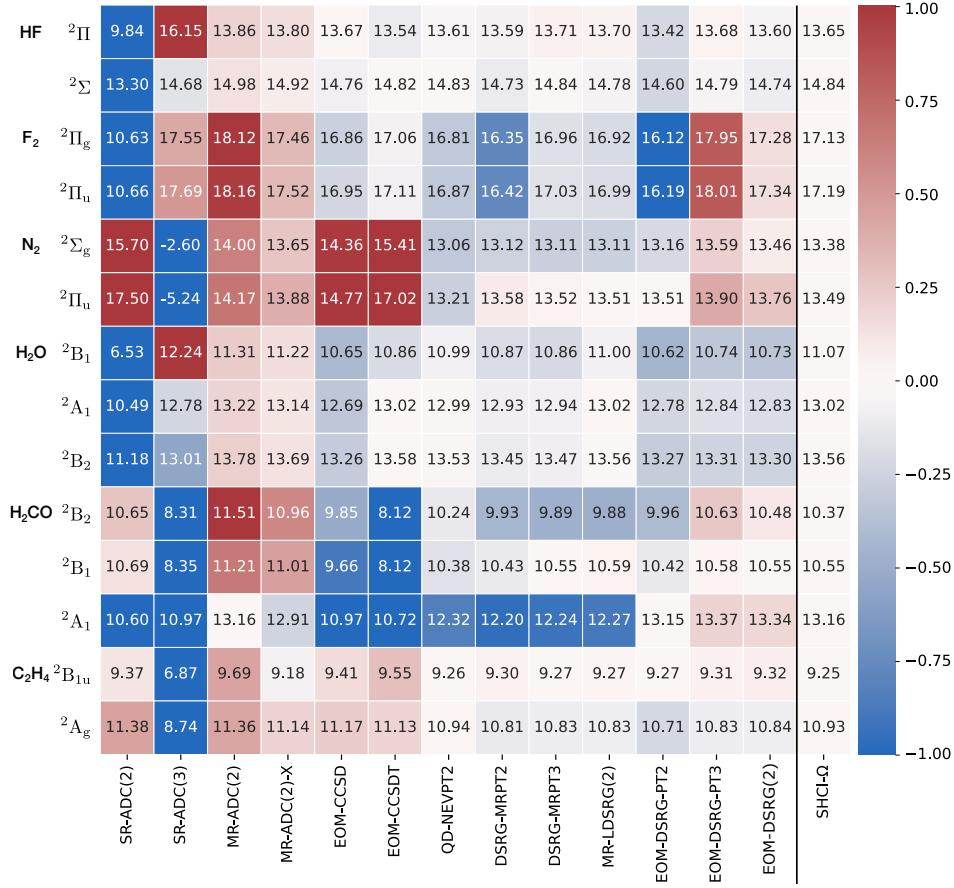

FIG. S2. Vertical ionization energies (IPs) of molecules at stretched geometries. The colors indicate the signed error (eV) of the IPs with respect to the SHCI values.

5

# IV. OVERALL FLOW PARAMETER DEPENDENCE OF THE EOM-DSRG IPS

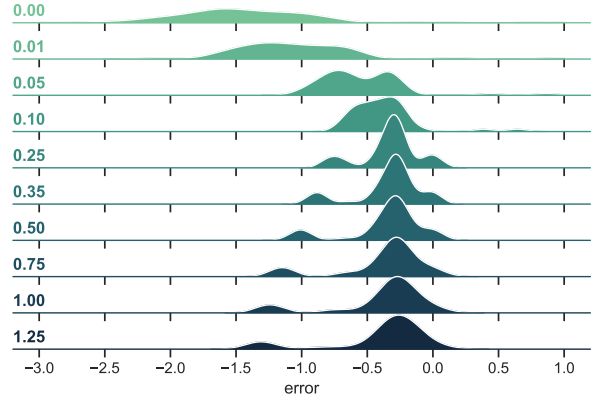

FIG. S3. Overall flow parameter dependence of the EOM-DSRG-PT2 IPs at both equilibrium and stretched geometries.

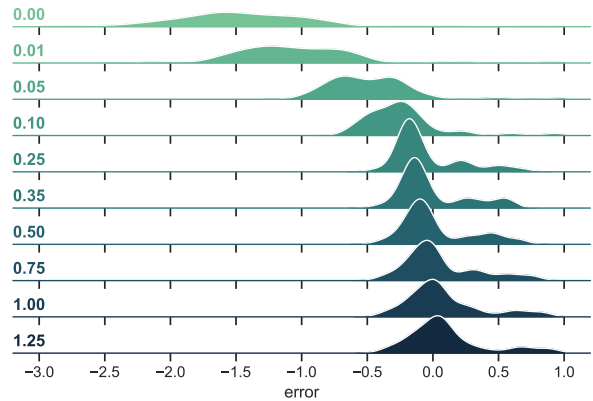

FIG. S4. Overall flow parameter dependence of the EOM-DSRG-PT3 IPs at both equilibrium and stretched geometries.

## V. FLOW PARAMETER DEPENDENCE OF THE EOM-DSRG IPS AT EQUILIBRIUM GEOMETRIES

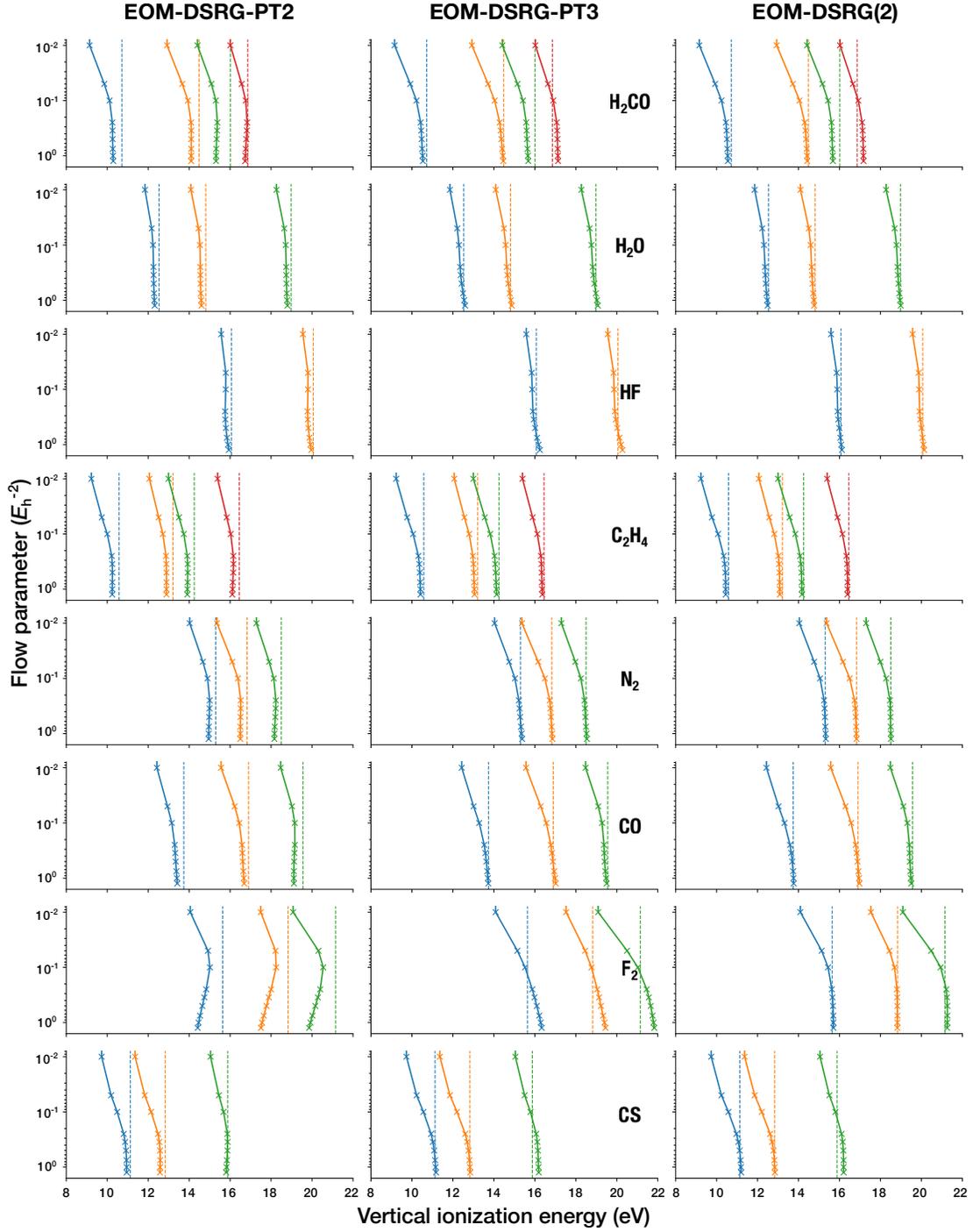

FIG. S5. Dependence of the ionization potentials (IPs) on the flow parameter.

# VI. FLOW PARAMETER DEPENDENCE OF THE EOM-DSRG IPS AT STRETCHED GEOMETRIES

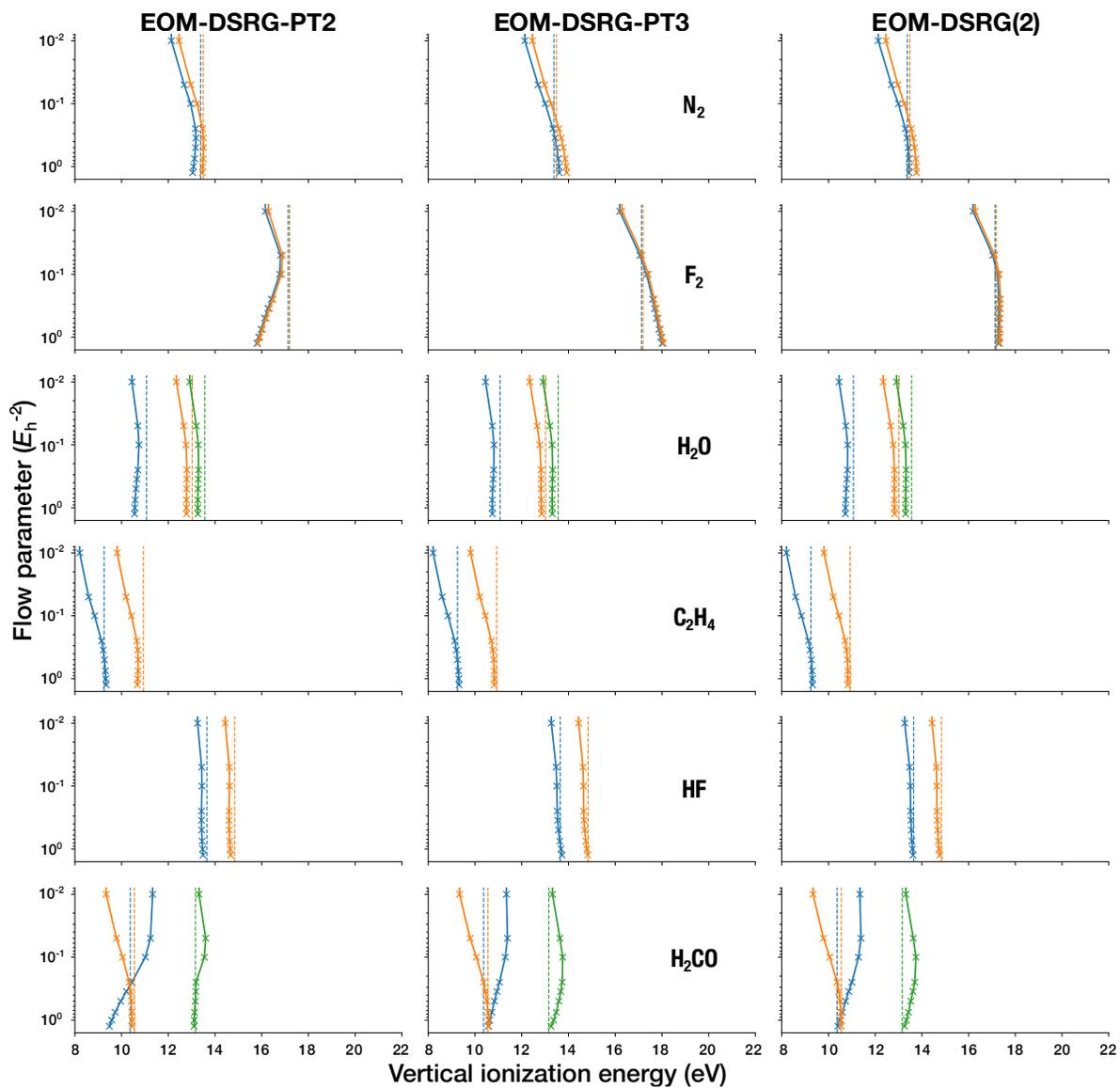

FIG. S6. Dependence of the ionization potentials (IPs) on the flow parameter.

# VII. SPECTROSCOPIC CONSTANTS OF SELECT RADICALS

## A. Equilibrium bond lengths

| Molecule | State | UHF-CCSD | UHF-CCSD(T) | EOMIP-CCSD | EOMIP-CCSD* | EOM-DSRG | Expt |
|---|---|---|---|---|---|---|---|
| OH | $X\ ^2\Pi$ | 0.9673 | 0.9694 | 0.9647 | 0.9688 | 0.9698 | 0.9697 |
| OH | $A\ ^2\Sigma^+$ | 1.0047 | 1.0061 | 0.9966 | 1.0045 | 1.0063 | 1.0121 |
| CN | $X\ ^2\Sigma^+$ | 1.1640 | 1.1707 | 1.1650 | 1.1710 | 1.1759 | 1.1718 |
| CN | $A\ ^2\Pi$ | 1.2284 | 1.2350 | 1.2221 | 1.2294 | 1.2383 | 1.2333 |
| CN | $B\ ^2\Sigma_b^+$ | 1.1368 | 1.1499 | 1.1415 | 1.1464 | 1.1547 | 1.150 |
| $N_2^+$ | $X\ ^2\Sigma_g^+$ | 1.1110 | 1.1192 | 1.1073 | 1.1132 | 1.1208 | 1.1164 |
| $N_2^+$ | $A\ ^2\Pi_u$ | 1.1674 | 1.1765 | 1.1570 | 1.1743 | 1.1806 | 1.1750 |
| $N_2^+$ | $B\ ^2\Sigma_u^+$ | 1.0623 | 1.0744 | 1.0647 | 1.0687 | 1.0778 | 1.0742 |
| $CO^+$ | $X\ ^2\Sigma^+$ | 1.1104 | 1.1162 | 1.1086 | 1.1170 | 1.1193 | 1.1151 |
| $CO^+$ | $A\ ^2\Pi$ | 1.2395 | 1.2471 | 1.2362 | 1.2336 | 1.2506 | 1.2437 |

TABLE S1. Comparison of of equilibrium bond lengths ($r_e$, Å) for select low-lying electronic states of various radicals. All theoretical values other than those from EOM-DSRG are from Saeh and Stanton,[2] experimental values are from Huber and Herzberg.[3]

## B. Harmonic vibrational frequencies

| Molecule | State | UHF-CCSD | UHF-CCSD(T) | EOMIP-CCSD | EOMIP-CCSD* | EOM-DSRG | Expt |
|---|---|---|---|---|---|---|---|
| OH | $X\ ^2\Pi$ | 3796 | 3763 | 3861 | 3763 | 3752 | 3738 |
| OH | $A\ ^2\Sigma^+$ | 3295 | 3282 | 3425 | 3300 | 3257 | 3179 |
| CN | $X\ ^2\Sigma^+$ | 2178 | 2134 | 2165 | 2115 | 2070 | 2069 |
| CN | $A\ ^2\Pi$ | 1877 | 1841 | 1927 | 1874 | 1809 | 1813 |
| CN | $B\ ^2\Sigma_b^+$ | 2375 | 2212 | 2339 | 2276 | 2188 | 2164 |
| $N_2^+$ | $X\ ^2\Sigma_g^+$ | 2292 | 2215 | 2352 | 2277 | 2200 | 2207 |
| $N_2^+$ | $A\ ^2\Pi_u$ | 2006 | 1922 | 2095 | 1924 | 1891 | 1904 |
| $N_2^+$ | $B\ ^2\Sigma_u^+$ | 2633 | 2486 | 2633 | 2582 | 2444 | 2420 |
| $CO^+$ | $X\ ^2\Sigma^+$ | 2313 | 2293 | 2315 | 2242 | 2220 | 2214 |
| $CO^+$ | $A\ ^2\Pi$ | 1626 | 1573 | 1651 | 1672 | 1551 | 1562 |

TABLE S2. Comparison of harmonic vibrational frequencies ($\omega_e$, cm$^{-1}$) for select low-lying electronic states of various radicals. All theoretical values other than those from EOM-DSRG are from Saeh and Stanton,[2] experimental values are from Huber and Herzberg.[3]

## C. Adiabatic electronic excitation energies

TABLE S3. Transition Energies for Various Electronic States

| Molecule | Transition | UHF-CCSD | UHF-CCSD(T) | EOMIP-CCSD | EOMIP-CCSD* | EOM-DSRG | Expt |
|---|---|---|---|---|---|---|---|
| OH | $X\ ^2\Pi \to A\ ^2\Sigma^+$ | 33561 | 33472 | 33709 | 33499 | 32602 | 32684 |
| CN | $X\ ^2\Sigma^+ \to A\ ^2\Pi$ | 7704 | 8695 | 9708 | 8831 | 9129 | 9245 |
| CN | $X\ ^2\Sigma^+ \to B\ ^2\Sigma^+$ | 26840 | 26865 | 28115 | 26262 | 25570 | 25752 |
| $N_2^+$ | $X\ ^2\Sigma_g^+ \to A\ ^2\Pi_u$ | 7198 | 8969 | 11228 | 8144 | 8782 | 9167 |
| $N_2^+$ | $X\ ^2\Sigma_g^+ \to B\ ^2\Sigma_u^+$ | 27082 | 26322 | 25490 | 26025 | 25383 | 25461 |
| CO$^+$ | $X\ ^2\Sigma^+ \to A\ ^2\Pi$ | 19281 | 20034 | 19454 | 21367 | 20529 | 20733 |

TABLE S4. Comparison of adiabatic electronic excitation energies ($T_e$, cm$^{-1}$) for select low-lying electronic states of various radicals. All theoretical values other than those from EOM-DSRG are from Saeh and Stanton,[2] experimental values are from Huber and Herzberg.[3]

11

## VIII. CN BINDING CURVE FOR PT2 AND PT3

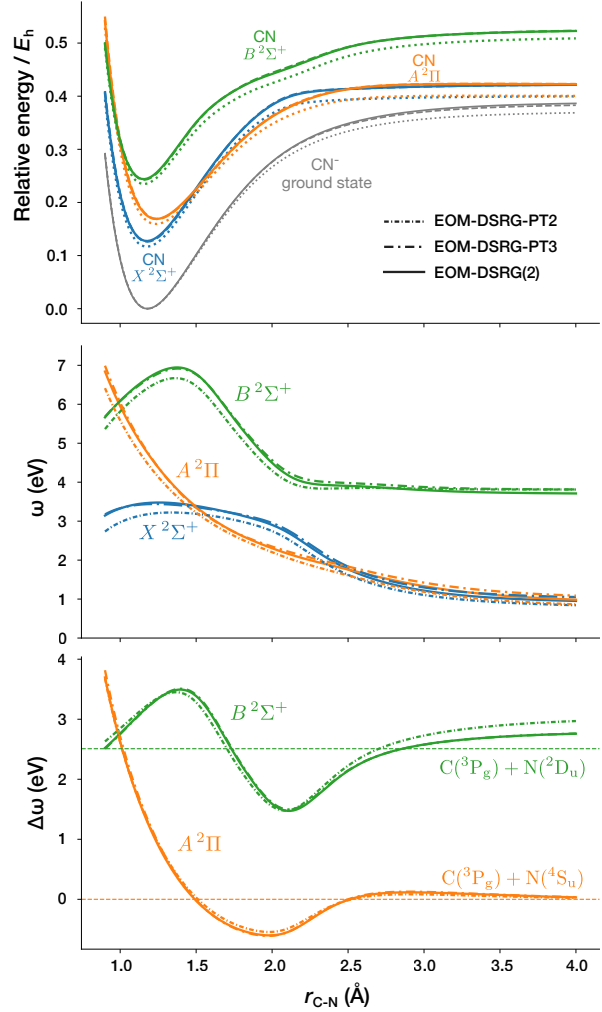

FIG. S7. Similar to Fig. 10 in the main text, but comparing IP-EOM-DSRG-PT2/3 to IP-EOM-DSRG(2).

# TOC Graphic

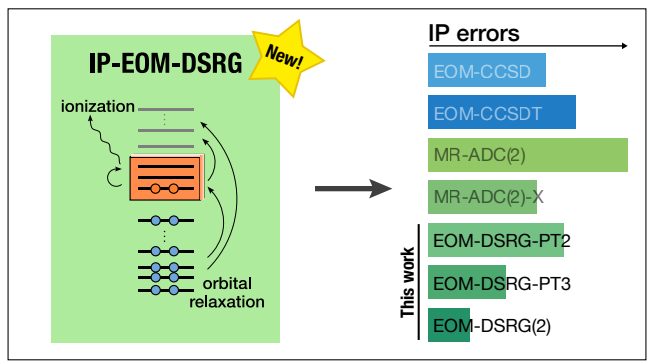



\end{document}